\documentclass[prb,twocolumn,showpacs,preprintnumbers,amsmath,amssymb]{revtex4}

\usepackage{amsmath,amssymb,amsfonts} 	% Typical maths resource package
\usepackage[dvips]{graphicx}
\usepackage[latin1]{inputenc}
\usepackage{subfigure}
\begin{document}

%\title{Accurate total energies of molecules and solids with the renormalized adiabatic local density approximation: From dispersive interactions to local correlation}
\title{Beyond the random phase approximation: Improved description of short range correlation by a renormalized adiabatic local density approximation}

\author{Thomas Olsen}
\email{tolsen@fysik.dtu.dk}
\author{Kristian S. Thygesen}
\email{thygesen@fysik.dtu.dk}

\affiliation{Center for Atomic-Scale Materials Design (CAMD) and Center for Nanostructured Graphene (CNG),
	     Department of Physics, Technical University of Denmark,
	     DK--2800 Kongens Lyngby, Denmark}

\date{\today}

\begin{abstract}
We assess the performance of a recently proposed renormalized adiabatic local density approximation (rALDA) for \textit{ab initio} calculations of electronic correlation energies in solids and molecules. The method is an extension of the random phase approximation (RPA) derived from time-dependent density functional theory and the adiabatic connection fluctuation-dissipation theorem and contains no fitted parameters. The new kernel is shown to preserve the accurate description of dispersive interactions from RPA while significantly improving the description of short range correlation in molecules, insulators, and metals. For molecular atomization energies the rALDA is a factor of 7(4) better than RPA(PBE) when compared to experiments, and a factor of 3(1.5) better than RPA(PBE) for cohesive energies of solids. For transition metals the inclusion of full shell semi-core states is found to be crucial for both RPA and rALDA calculations and can improve the cohesive energies by up to 0.4 eV. Finally we discuss straightforward generalizations of the method, which might improve results even further.
\end{abstract}
\maketitle

\section{Introduction}
The adiabatic-connection fluctuation-dissipation theorem (ACFDT) provides an exact representation of the electronic correlation energy in term of the interacting density response function, within density functional theory (DFT).\cite{langreth,gunnarsson} A major advantage of this method, is that it naturally accounts for dispersive interactions through the non-locality of the response function. Furthermore, in contrast to semi-local approximations, the ACFDT correlation energy is naturally combined with the exact exchange energy and does not rely on error cancellation between the exchange and correlation contributions to the total energy. The accuracy of correlation energies within the ACFDT, then depends on the quality of the interacting response function which needs to be approximated.

The most famous approximation for the response function is the random phase approximation (RPA), which is obtained when a non-interacting approximation is used for the irreducible polarizability. For metallic systems, the RPA provides a qualitative account of screening in and cures the pathological divergence of second order perturbation theory for the homogeneous electron gas. In 2001 Furche\cite{furche} applied RPA and ACFDT to obtain the dissociation energies of small molecules and found that the results were slightly worse than those obtained with a generalized gradient approximation\cite{pbe} (GGA) with a systematic tendency to underbind. It was also shown that RPA can account for strong static correlation and correctly reproduces the dissociation limit of the N$_2$ molecule. Following this, RPA has been applied to calculate cohesive energies of solids\cite{harl10, olsen_rpa2, jun_rpa} and again, RPA performs significantly worse than GGA with a systematic tendency to underbind. In contrast, RPA produces excellent results for van der Waals bonded systems like graphite\cite{lebegue}, which is very poorly described by semi-local approximations. In addition, for graphene adsorbed on metal surfaces, where both covalent and dispersive interactions are equally important, the RPA seems to be the only non-fitted scheme capable of describing the potential energy curves correctly.\cite{olsen_rpa1,mittendorfer,olsen_rpa2}

By now, it is well established that RPA provides a reliable account of van der Waals bonded systems but systematically underestimates the strength of covalent and ionic bonds.\cite{eshuis, ren_review} Furthermore, the absolute correlation energies obtained with RPA are severely underestimated and dissociation energies benefit from huge error cancellations. In particular, for one-electron systems RPA gives rise to a substantial negative correlation energy. This large self-correlation error can be remedied by subtracting the local RPA error obtained from the homogeneous electron gas,\cite{yan} but unfortunately the procedure does not improve upon dissociation energies of molecules and solids.\cite{furche,harl10} A more sophisticated approach is to add a second order screened exchange (SOSEX) contribution to the correlation energy, which exactly cancels the self-correlation energy for one-electron systems. This approach has been shown to improve dissociation energies of molecules\cite{ren_review} and cohesive energies of solids\cite{gruneis}, but is significantly more computationally demanding than RPA. In addition the SOSEX term in the correlation energy destroys the good description of static correlation in RPA and produces the wrong dissociation limit of small molecules.\cite{ren_review}

In a different line of development, time-dependent density functional theory\cite{runge-gross} (TDDFT) provides a systematic way to improve the RPA response function. Here the response function can be expressed in terms of a frequency-dependent non-local exchange-correlation kernel and RPA is obtained when the kernel is neglected. A rather advanced approach in this direction, is the inclusion of the frequency-dependent exact exchange kernel, which has been shown to produce very accurate dissociation energies of small molecules\cite{hellgren,hesselmann1} and conserve the accurate description of static correlation characteristic of RPA.\cite{hesselmann2} While this method is considerably more involved than RPA, it provides evidence that accurate correlation energies may be obtained from TDDFT and ACFDT with a good approximation for the exchange kernel. In Ref. [\onlinecite{lein}] the correlation energy of the homogeneous electron gas was evaluated using different approximations for the exchange-correlation kernel and the results indicated that the frequency dependence of the kernel is of minor importance, while the non-locality of the kernel is crucial. Moreover, it has been shown by Furche and van Voorhis\cite{furche_voorhis} that any local approximation for the kernel produces a correlation hole, which diverges at the origin. The resulting correlation energies then often become worse than those obtained with RPA (one exception to this is the local energy-optimized kernel of Ref. \cite{dobson_wang00}). Whereas exchange-correlation kernels have traditionally been derived to produce accurate excited state properties, there is now a considerable interest in obtaining exchange-correlation kernels suited for accurate ground state correlation energies.\cite{dobson_wang00, jung, gould} In this respect, it is interesting to note that the optical properties of electronic systems are ill described with local approximations for the kernel due to wrong behavior at $q\rightarrow0$, while the failure for total energy calculations originate from the bad behavior in the limit $q\rightarrow\infty$ (see discussion below).

In this paper we present a parameter-free renormalized adiabatic exchange kernel. The renormalization introduces non-locality in the kernel and provides an accurate description of the correlation hole at short distances, which gives rise to a better description of short-range correlation compared to RPA. We note that the philosophy of the renormalization is similar to the smooth cutoff introduced in the energy-optimized kernel of Ref. \cite{jung}. However, in contrast to that kernel, the present kernel does not contain any fitted parameters. The kernel has previously been shown to improve upon RPA for absolute correlation energies and dissociation energies of small molecules\cite{olsen_ralda1}, while the computational load is comparable to RPA. Here we describe the theory and implementation in detail and assess the performance for cohesive energies of solids, static correlation and van der Waals interactions.

\section{Theory}
Using the adiabatic connection and fluctuation-dissipation theorem (ACFDT), the exchange-correlation energy can be written as:
\begin{align}\label{E_xc}
 E_{xc}=-\int_0^1d\lambda\int_0^\infty\frac{d\omega}{2\pi}\text{Tr}\Big\{v[\hat{n}2\pi\delta(\omega)+\chi^{\lambda}(i\omega)]\Big\},
\end{align}
where $\hat{n}(\mathbf{r},\mathbf{r}')=n(\mathbf{r})\delta(\mathbf{r}-\mathbf{r}')$ and $v$ is the Coulomb interaction. Here $n(\mathbf{r})$ is the density, which by definition is constant along the adiabatic connection and $\chi^\lambda(i\omega)$ is the interacting response function of a system with $v\rightarrow\lambda v$ evaluated at imaginary frequencies. It is standard practice to divide $E_{xc}$ into an exchange part $E_x$ obtained by setting $\lambda=0$ in the integrand and a correlation part $E_c$, which is the remainder. One then obtains 
\begin{align}
 E_{x}&=-\int_0^\infty\frac{d\omega}{2\pi}\text{Tr}\Big\{v[\hat{n}2\pi\delta(\omega)+\chi^{KS}(i\omega)]\Big\},\label{E_x}\\
E_{c}&=-\int_0^1d\lambda\int_0^\infty\frac{d\omega}{2\pi}\text{Tr}\Big\{v[\chi^\lambda(i\omega)-\chi^{KS}(i\omega)]\Big\},\label{E_c}
\end{align}
where $\chi^{KS}(i\omega)$ is the response function of the non-interacting Kohn-Sham system. A major advantage of this separation is that the exchange energy only depends on the occupied bands and can be converged separately with respect to $k$-points plane wave cutoff, etc. One can then focus on the correlation energy $E_c$, which can be calculated once an approximation for $\chi^\lambda$ has been given.

To obtain $\chi^\lambda$ we turn to time-dependent density functional theory, where it is easy to show that the interacting response function satisfies the Dyson equation
\begin{align}\label{dyson}
\chi^\lambda(\omega)=\chi^{KS}(\omega)+\chi^{KS}(\omega)\bigg\{\Big[\lambda v+f^\lambda_{xc}(\omega)\Big]\chi^\lambda(\omega)\bigg\}.
\end{align}
Here, the exchange-correlation kernel $f_{xc}^\lambda(\omega)$ is the temporal Fourier transform of the functional derivative of the time-dependent exchange-correlation potential at coupling strength $\lambda$. All the complicated correlation effects contained in $\chi^\lambda(\omega)$ has been transferred into $f_{xc}^\lambda(\omega)$, which needs to be approximated. However, even if $f_{xc}^\lambda(\omega)$ is neglected all together, one still obtains a non-trivial approximation for $\chi^\lambda(\omega)$ due to the Coulomb interaction term. This is the random phase approximation (RPA). 

To obtain correlation energies beyond the random phase approximation, we will include an approximation for $f_{xc}^\lambda(\omega)$ in Eq. \eqref{dyson}. In general, one can obtain the exchange-correlation kernel along the adiabatic connection from the scaling properties \cite{lein}
\begin{align}
f^\lambda_{xc}[n](\mathbf{r},\mathbf{r'},\omega)&=\lambda^2f_{xc}[n'](\lambda\mathbf{r},\lambda\mathbf{r'},\omega/\lambda^2),\\
n'&=\lambda^{-3}n(\mathbf{r}/\lambda),
\end{align}
and it will thus be sufficient to consider the case of $\lambda=1$. Due to the first order nature of exchange, any properly derived pure exchange kernel should have the property that $f^\lambda_{x}[n](\mathbf{r},\mathbf{r'},\omega)=\lambda f_{x}[n](\mathbf{r},\mathbf{r'},\omega)$. For a pure exchange kernel the coupling constant integration can thus be carried out resulting in
\begin{align}
E_{c}=\int_0^\infty\frac{d\omega}{2\pi}\text{Tr}\Big\{vf_{Hx}^{-1}(i\omega)\ln&[1-\chi^{KS}(i\omega)f_{Hx}(i\omega)]\notag\\
&+v\chi^{KS}(i\omega)]\Big\},\label{E_c_int}
\end{align}
where $f_{Hx}(i\omega)=v+f_x(i\omega)$. However, the inversion of $f_{Hx}$ turns out to cause numerical problems and for \textit{ab initio} applications in this manuscript we will perform the coupling constant integration numerically. 

\subsection{Adiabatic Local Density Approximation}
A simple and natural choice is the adiabatic local density approximation (ALDA) given in the frequency domain by 
\begin{align}
f^{\text{ALDA}}_{xc}[n](\mathbf{r},\mathbf{r'})&=\delta(\mathbf{r}-\mathbf{r'})f^{\text{ALDA}}_{xc}[n(\mathbf{r})],\label{ALDA1}\\
f^{\text{ALDA}}_{xc}[n(\mathbf{r})]&=\frac{d^2}{dn^2}\Big(n\varepsilon_{xc}^{HEG}\Big)\Big|_{n=n(\mathbf{r})},\label{ALDA2}
\end{align}
where $\varepsilon_{xc}^{HEG}$ is the exchange-correlation energy per electron in the homogeneous electron gas. The approximation is in a certain sense similar to LDA in static DFT, but in contrast to LDA, the ALDA is not exact for the homogeneous electron gas. In particular, the ALDA kernel becomes a (density dependent) constant in the homogeneous electron gas, while the true kernel should depend on both frequencies and position differences. This means that ALDA can violate a number of exact conditions. For example, it is well known that the kernel in Fourier space should behave as $f_{xc}\sim q^{-2}$ for $q\rightarrow 0$ and $f_{xc}\sim q^{-2}$ for $q\rightarrow\infty$, \cite{onida} which are obviously violated by any local kernel. Whereas, the first of these conditions is very important for the description of optical excitations within TDDFT, it is the second condition, which makes any local approximation $f_{xc}$ useless for total energy calculations in the ACFDT framework. The reason is that the trace in Eq. \eqref{E_c} becomes an integral over all $\mathbf{q}$ and an inaccurate description of the response function at large $q$, can deteriorate the correlation energy completely. Equivalently, the constant behavior of $f_{xc}$ at large $q$ renders the pair correlation function in real-space divergent at the origin. This divergence is integrable, but gives rise to severe convergence problems for \textit{ab initio} applications.\cite{furche_voorhis}

\begin{figure}[tb]
	\includegraphics[width=4.25 cm]{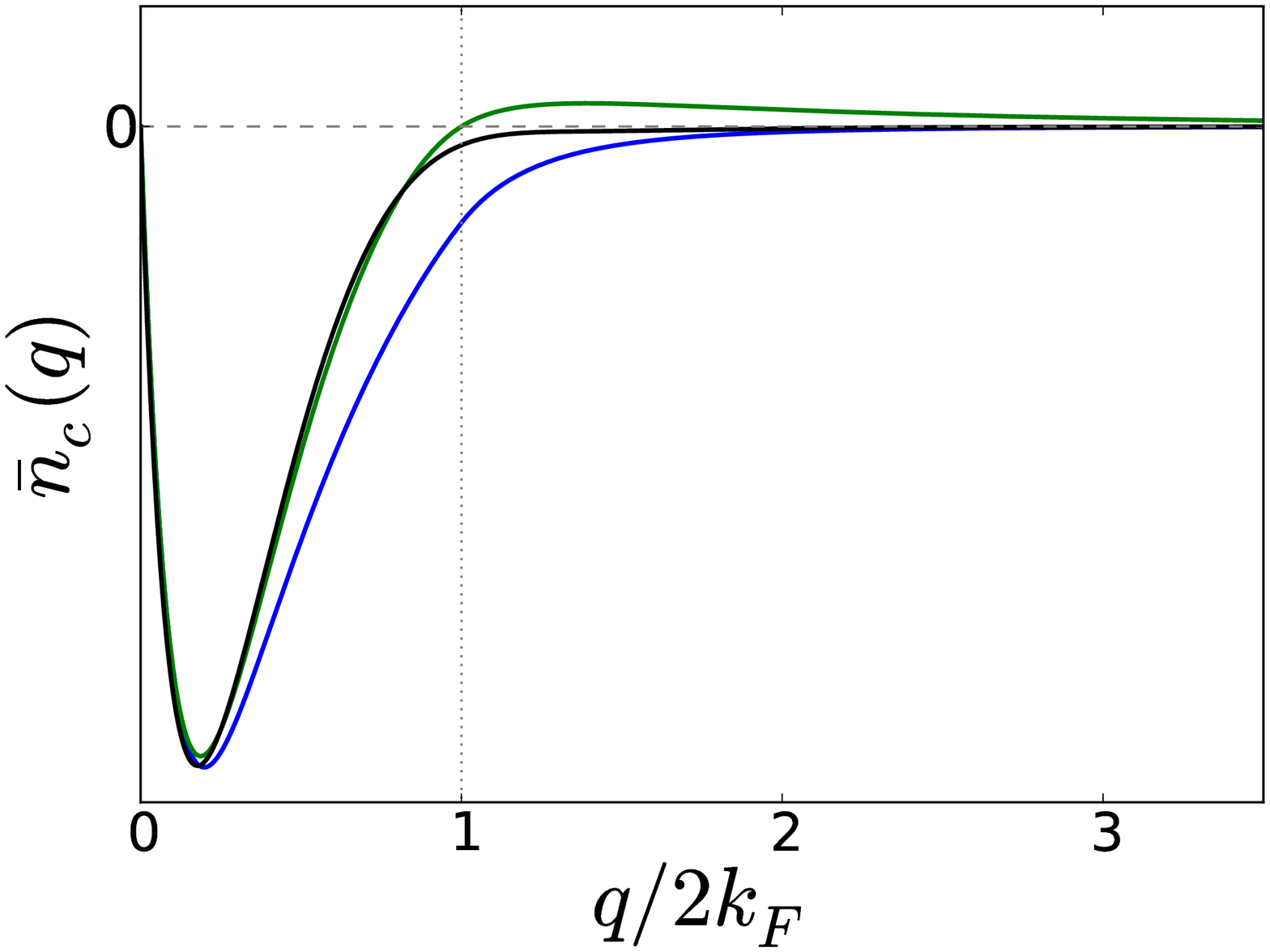} 
        \includegraphics[width=4.25 cm]{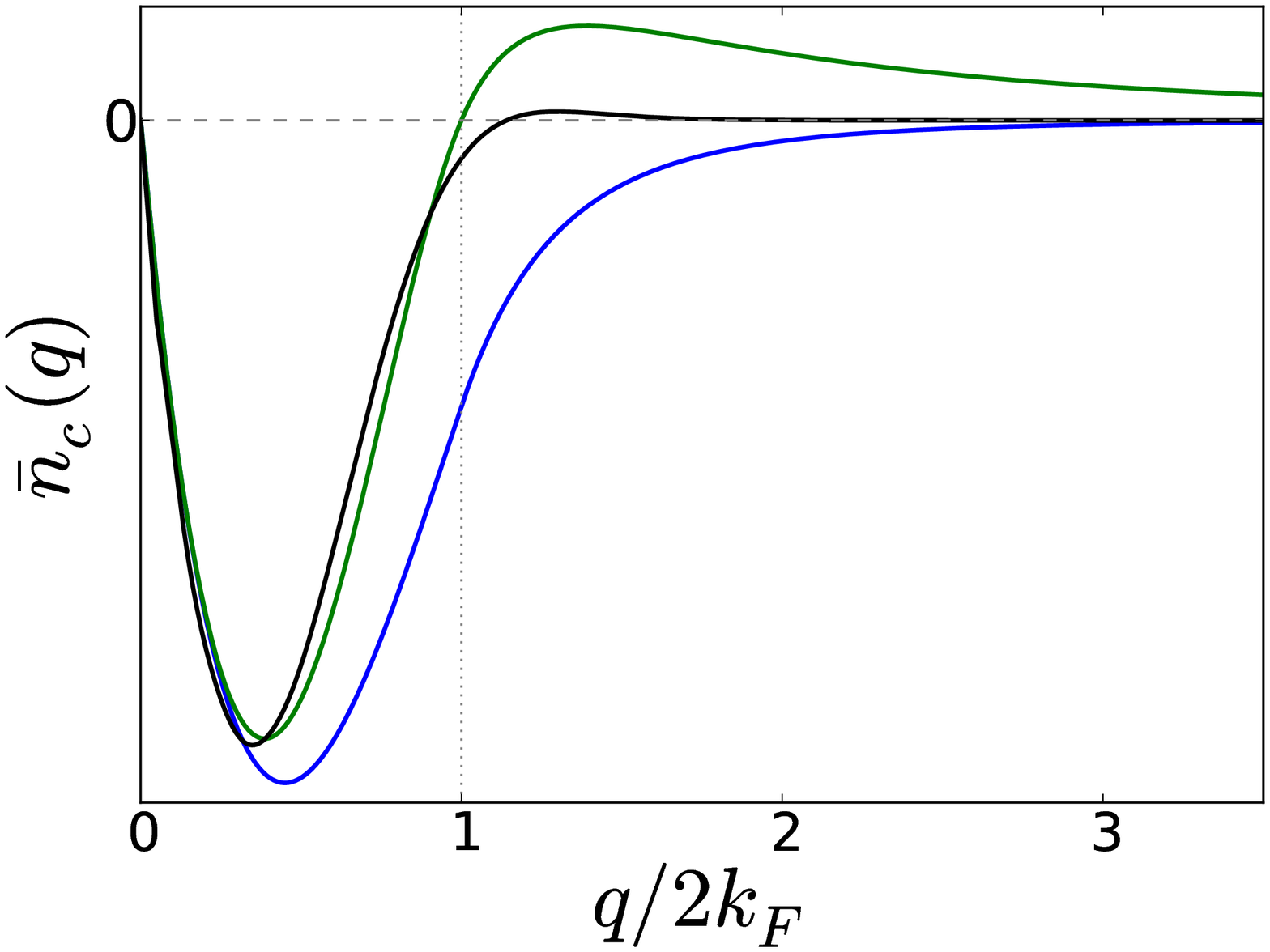}
\caption{(color online). The coupling constant averaged correlation hole in $q$-space for the homogeneous electrons gas. Left: $r_s=1$. Right: $r_s=10$. The positive slowly decaying tail of ALDA is evident at large $q$. The rALDA$_X$ is obtained by truncating $\hat n_c^{ALDA_X}(q)$ at its zero point at $2k_F.$}
\label{fig:g_q}
\end{figure}

To show this in more detail we have plotted the correlation hole of the homogeneous electron gas in the top row of Fig. \ref{fig:g_q} using RPA and the exchange part of the ALDA (ALDA$_X$) approximation for $f_{xc}$ and compare with an analytic representation.\cite{perdew_wang} The analytic representation has been shown to agree with Monte Carlo simulations and here we will regard this as the exact correlation hole. The correlation energy per electron is directly related to the integral of the coupling constant averaged correlation hole $\bar{n}_c$:
\begin{equation}
 E_c=2\pi\int_0^\infty dr r\bar{n}_c(r)=\frac{1}{\pi}\int_0^\infty dq \bar{n}_c(q).
\end{equation}
It is most instructive to look at the correlation hole in $q$-space. It is seen that the RPA hole decays to slowly and since the correlation energy is proportional to the integral of the correlation hole, is is clear that RPA will overestimate the magnitude of the correlation energy. In fact, the correlation energy per electron becomes $\sim0.5$ eV too negative for a wide range of densities, whereas the relative error increases from $~25\%$ at $r_s=1$ to $~50\%$ at $r_s=20$.\cite{olsen_ralda1}. ALDA$_X$, on the other hand, seems to reproduce the $q$-space representation of the correlation hole much better. However, at large $q$, the correlation hole acquires a slowly decaying positive tail and as a consequence, the correlation energy becomes too large by $\sim0.3$ eV per electron. Furthermore, the slowly decaying tail introduces a divergence at the origin in the real space correlation hole. We note that the results change very little if we include the full ALDA (exchange + correlation) kernel.

\begin{figure}[tb]
	\includegraphics[width=4.25 cm]{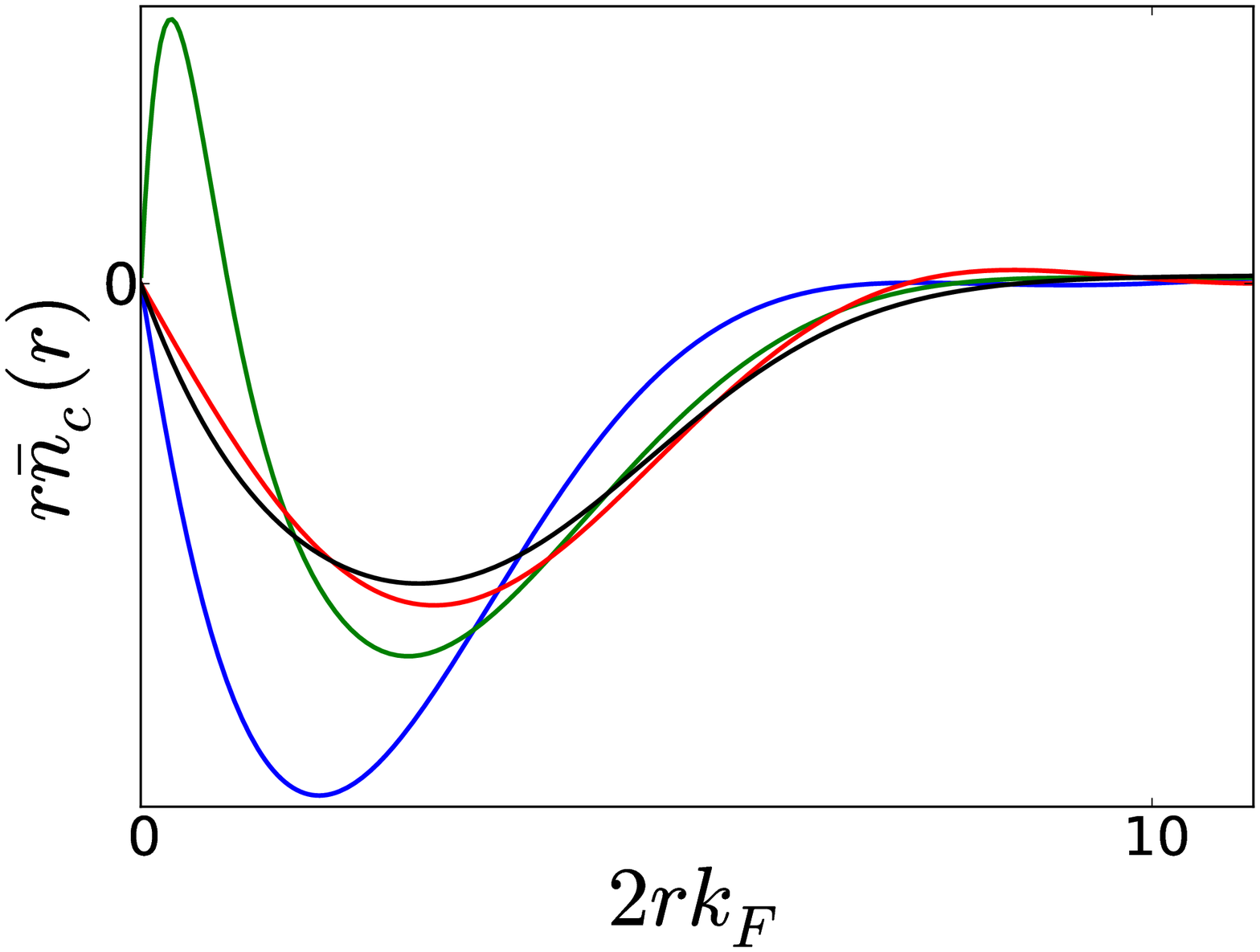} 
        \includegraphics[width=4.25 cm]{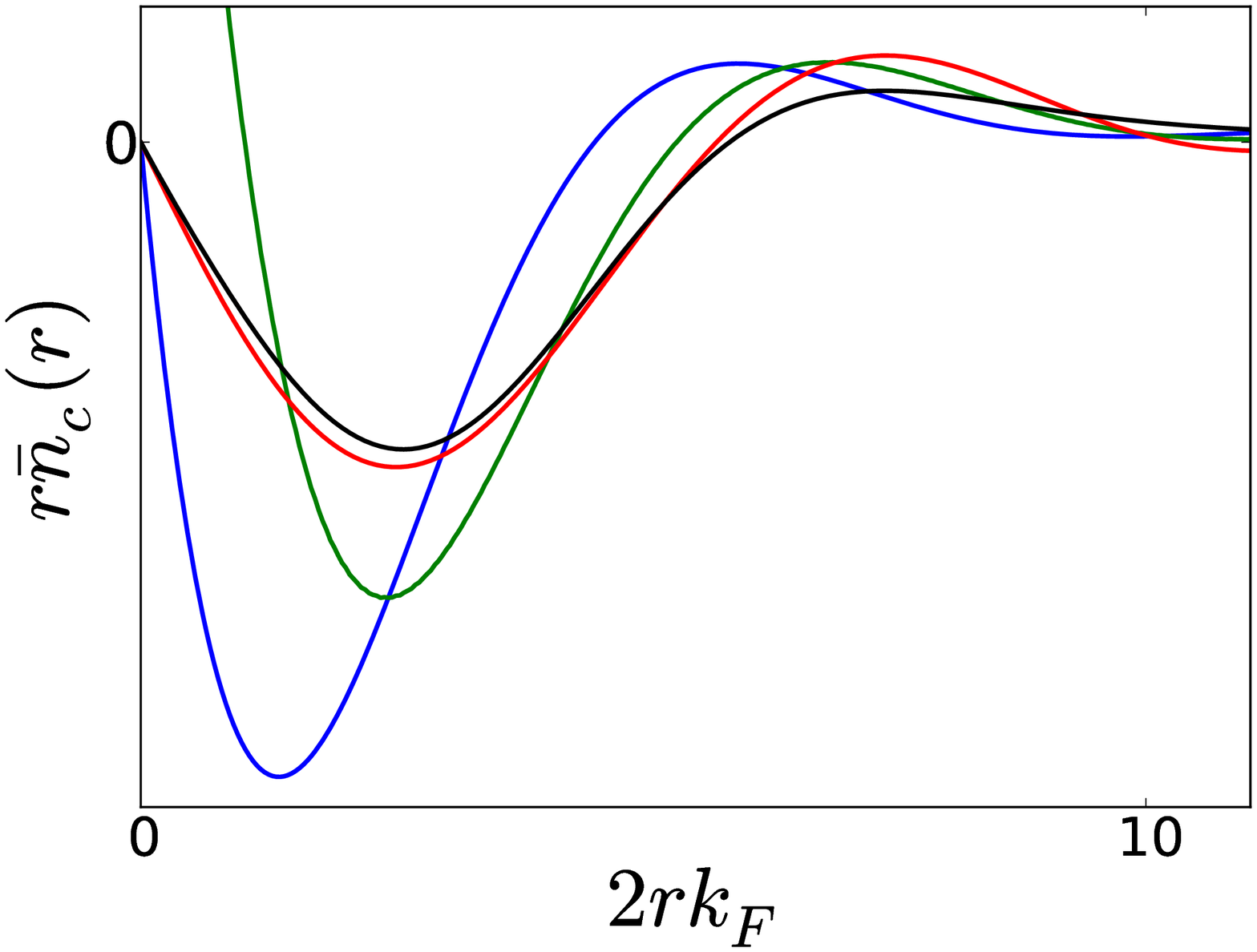}\\
	\includegraphics[width=4.25 cm]{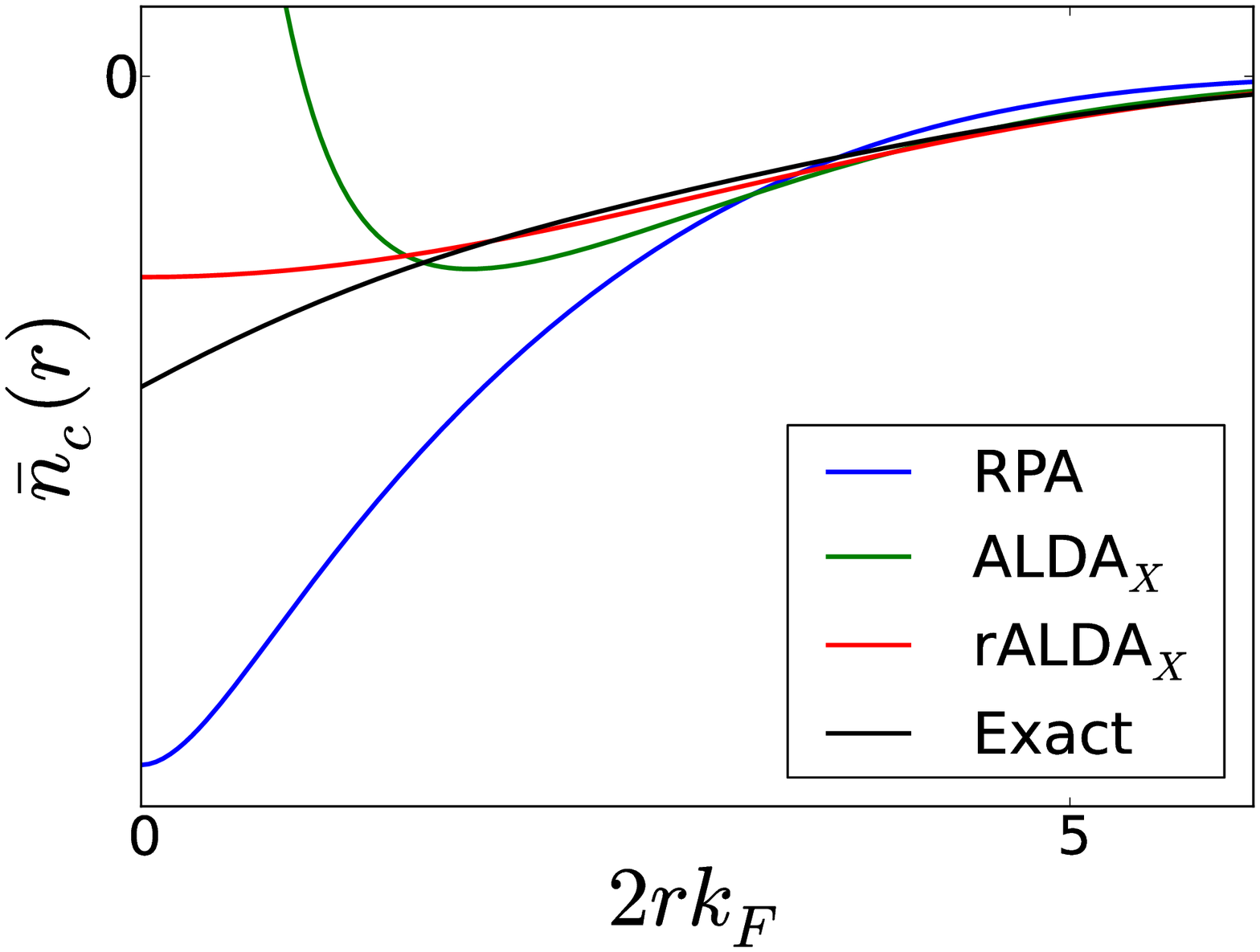} 
        \includegraphics[width=4.25 cm]{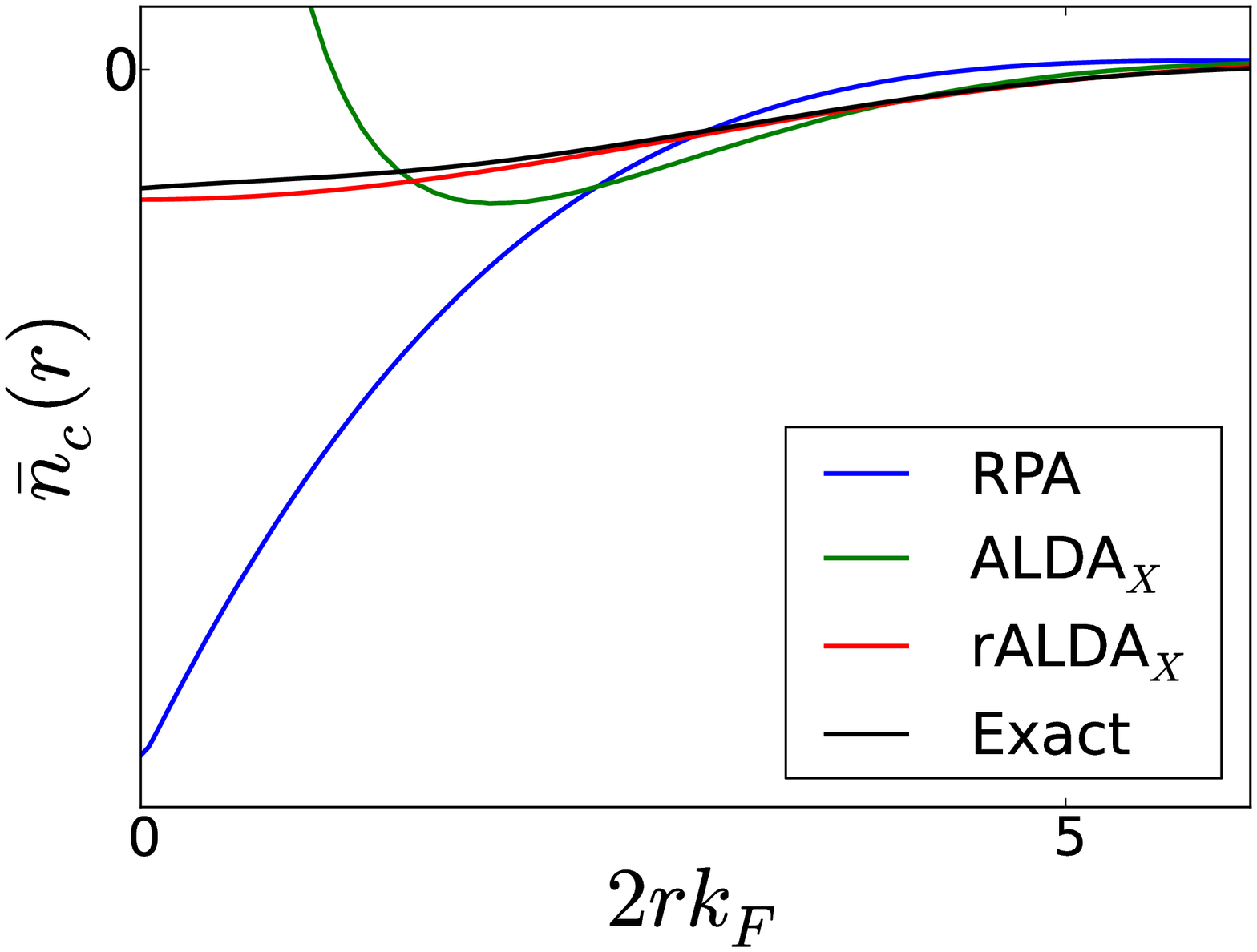}	
\caption{(color online). The coupling constant averaged correlation hole in real-space for the homogeneous electrons gas. Left column: $r_s=1$. Right column: $r_s=10$. The bottom row shows the bare correlation hole in real space, where the ALDA$_X$ approximation is seen to diverge at the origin. The rALDA$_X$ is obtained by truncating $\bar n_c^{ALDA_X}(q)$ at $2k_F$ (the zero point of $\bar n_c^{ALDA_X}(q)$) and is seen to produce a much better approximation than both RPA and ALDA$_X$. The top row, shows the correlation hole weighted by $r$, which is the quantity one would integrate to get the correlation energy.}
\label{fig:g_r}
\end{figure}

If we are mostly concerned with total correlation energies, we may observe that the exact correlation hole picks up most of its weight between zero and $2k_F$ where $\bar n_c^{\text{ALDA}_X}(q)$ has a zero-point. The correlation energy is then approximated by $E_c\propto\int_0^{2k_F}\bar n_c^{\text{ALDA}_X}(q)dq$ and we have previously shown that this procedure yields correlation energies within 30 meV of the exact value for a wide range of densities \cite{olsen_ralda1}. We can now proceed to define a new real space correlation hole obtained by Fourier transforming $\bar n_c^{\text{ALDA}_X}(q)$ truncated at $2k_F$ and we will refer to this as rALDA$_X$. From Fig. \ref{fig:g_r} it is evident that this procedure removes the divergence at the origin and gives rise to a correlation hole, which comprises a much better short range approximation than either RPA or ALDA$_X$. At the same time the rALDA$_X$ correlation hole retains the accurate long-range behavior of the ALDA$_X$ correlation hole, which is less accurately described in RPA. In Fig. \ref{fig:g_r} we also show the correlation hole weighted by $r$, which is the quantity that should be integrated to obtain the correlation energy.

\subsection{Renormalized Adiabatic Local Density Approximation}
\begin{figure}[tb]
	\includegraphics[width=4.25 cm]{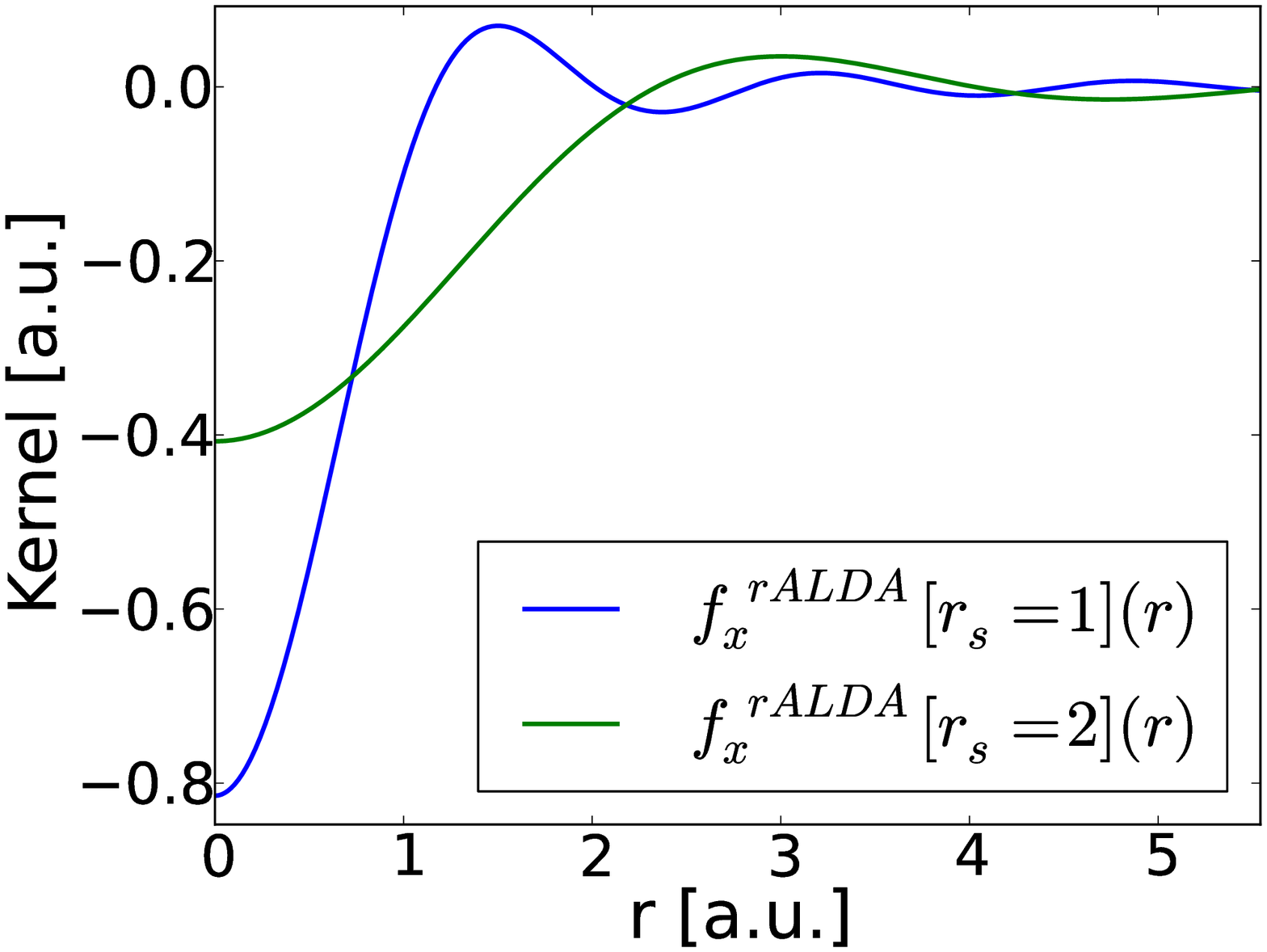} 
        \includegraphics[width=4.25 cm]{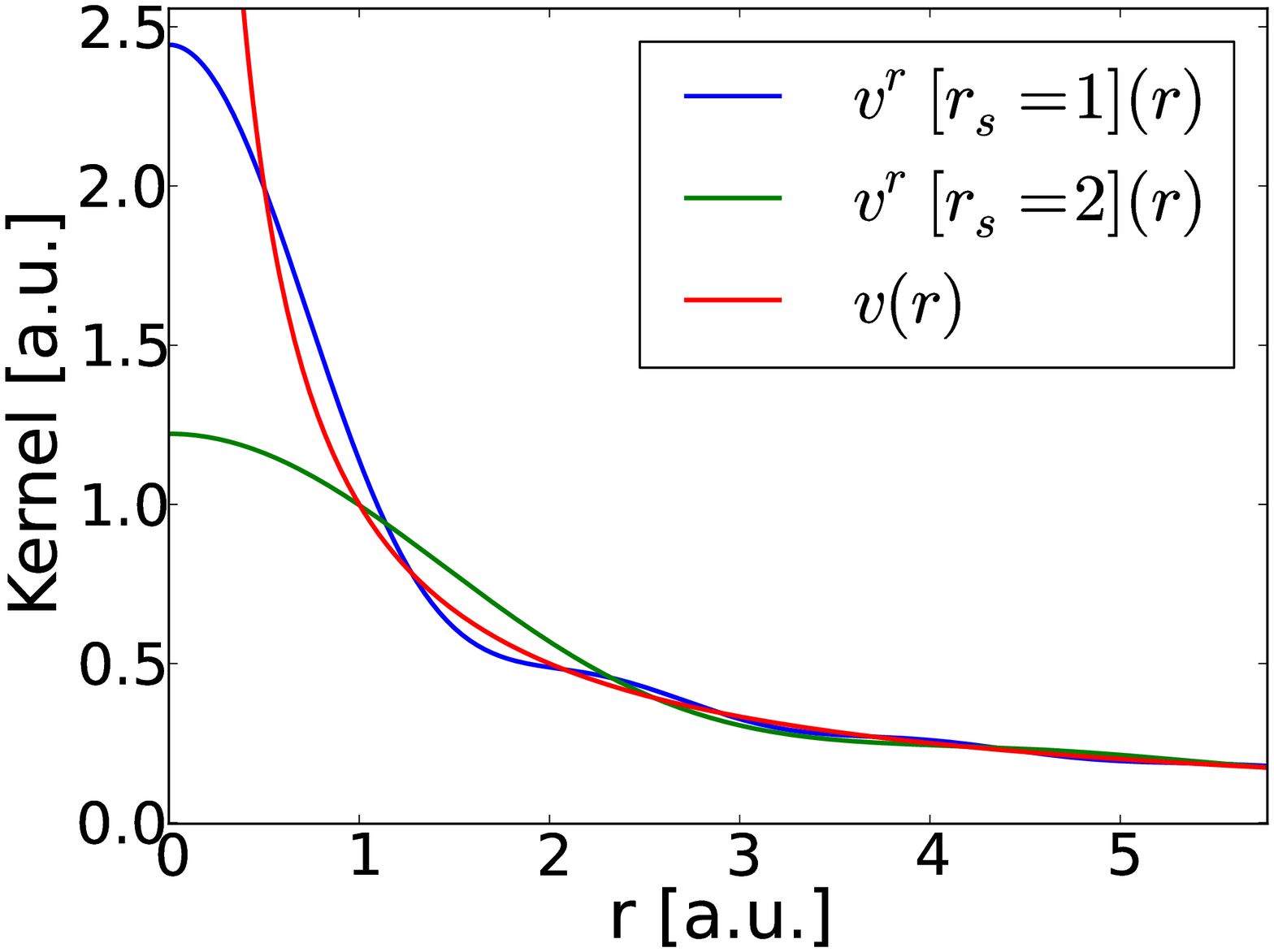}	
\caption{(color online). Left: The renormalized exchange kernel $\widetilde{f}_x^{rALDA}[n](r)$. The kernel has a width which is determined by the density. As $n\rightarrow\infty$ ($r_s\rightarrow0$) the width decreases and the kernel approaches the bare exchange kernel \eqref{ALDA1}. Right: The renormalized Hartree kernel $v_c^{rALDA}[n](r)$. The kernel is finite at the origin and approaches the bare Coulomb kernel for large $r$.}
\label{fig:rALDA}
\end{figure}

For the homogeneous electron gas, the cutoff at $2k_F$ in $\bar n_c^{ALDA_X}(q)$ can be imposed by using the Hartree-exchange kernel
\begin{align}
f_{Hx}^{rALDA_X}[n](q)=\theta\Big(2k_F-q\Big)f_{Hx}^{ALDA_X}[n].
\end{align}
Fourier transforming this expression yields
\begin{align}\label{rALDA}
f_{Hx}^{rALDA}[n](r)&=\widetilde{f}_{x}^{rALDA}[n](r)+v^r[n](r),\\ 
\widetilde{f}_{x}^{rALDA}[n](r)&=\frac{f^{ALDA}_{x}[n]}{2\pi^2r^3}\Big[\sin(2k_Fr)-2k_Fr\cos(2k_Fr)\Big],\notag\\
v^r[n](r)&=\frac{1}{r}\frac{2}{\pi}\int_0^{2k_Fr}\frac{\sin x}{x}dx,\notag
\end{align}
where $k_F=(3\pi^2n)^{1/3}$ and we have suppressed the ALDA exchange ($X$) index. We will refer to this kernel as the renormalized adiabatic local density approximation (rALDA). The cutoff in $q$-space is translated into a density dependent width of the delta function in Eq. \eqref{ALDA1}, which gives rise to a non-local exchange-correlation kernel. Similarly, the renormalized Hartree kernel acquires a density dependence through the cutoff, but approaches the bare Hartree kernel in the limit of $r\rightarrow\infty$. The kernels are shown in Fig. \ref{fig:rALDA}. 
Interestingly, both the renormalized Hartree kernel and exchange kernel become finite at the origin giving
\begin{align}
v^r[n](r\rightarrow0)&=\frac{4k_F}{\pi}-\frac{8k_F^3r^2}{9\pi},\label{kernel_origin}\\
\widetilde{f}_{x}^{rALDA}[n](r\rightarrow0)&=\Big[\frac{4k_F^3}{3\pi^2}-\frac{32k_F^5r^2}{15\pi^2}\Big]f^{ALDA}_{x}[n].
\end{align}
This property becomes extremely convenient when the kernel is evaluated in real space.

It is customary to include an exchange-correlation kernel on top of the exact Hartree kernel and we thus define the renormalized adiabatic local density approximation by the exchange kernel
\begin{align}\label{rALDA_exchange}
f_{x}^{rALDA}[n](r)&=\widetilde{f}_{x}^{rALDA}[n](r)+v^r[n](r)-v(r).
\end{align}
This representation is also more useful for \textit{ab initio} applications to solid state systems since it is difficult to converge the long range tail of $v^r[n](r)$. In contrast, for $r\rightarrow\infty$, [$v^r[n](r)-v(r)]\rightarrow\sin(2k_Fr)/r$, which rapidly averages to zero. Since $\widetilde{f}_{x}^{rALDA}[n](r)$ decays as $1/r^3$ for $r\rightarrow\infty$, it is much easier to converge the numerical Fourier transform of Eq. \eqref{rALDA_exchange} with respect to sampled unit cells than Eq. \eqref{rALDA}.

\subsubsection{Spin}
A major advantage of the RPA for \textit{ab initio} calculations of total correlation energies, is the fact the spin-polarized systems can be treated by simply making the substitution $\chi^0\rightarrow \chi^0_\uparrow+\chi^0_\downarrow$ in Eq. \eqref{dyson}. This is easily shown by using the fact that $f_{Hxc}$ is independent of spin in RPA, but it no longer holds when a spin-dependent exchange-correlation kernel is used.

For exact exchange, one has
\begin{align}\label{Exs}
E_x[n_\uparrow,n_\downarrow]=\frac{E_x[2n_\uparrow]+E_x[2n_\downarrow]}{2},
\end{align}
which translates into
\begin{align}\label{fxs}
f_{x,\sigma\sigma'}[n_\uparrow,n_\downarrow]=2f_x[2n_\sigma]\delta_{\sigma\sigma'}.
\end{align}
Here functionals of two arguments are the spinpolarized versions of the spinpaired functionals with one argument. It is straightforward to impose this spin-scaling on $f_x^{rALDA}$ as well, however, we find that this renders the correlation energy very difficult to converge, since the spin-polarized version of the kernel will inherit part of the convergence problems from ALDA. This is due to the fact that the spin-diagonal components of the Hartree-exchange kernel becomes $f_{Hx,\sigma\sigma}^{rALDA}=2f_x[2n_\sigma]+2v^r[2n_\sigma]-v$ and the additional bare Coulomb interaction $v$ destroys the local cancellation of the correlation hole resulting from the spin-density $n_\sigma$. 

To obtain the a more useful spin-polarized version of the rALDA kernel we first consider the Dyson equation with explicit spin dependence:
\begin{align}\label{dyson1}
\chi_{\sigma\sigma'}=\chi^{KS}_{\sigma}\delta_{\sigma\sigma'}+\sum_{\sigma''}\chi^{KS}_{\sigma}f^{Hxc}_{\sigma\sigma''}[n_\uparrow,n_\downarrow]\chi_{\sigma''\sigma'},
\end{align}
where we used that the non-interacting response function is diagonal in spin. For a spin-paired system, we get the constraint that
\begin{align}\label{dyson_spin_condition}
\frac{1}{4}\sum_{\sigma\sigma'}f^{xc}_{\sigma\sigma'}[n/2,n/2]=f_x[n].
\end{align}
Clearly this is satisfied if we impose the exact condition Eq. \eqref{fxs}. Due to the convergence problems mentioned above, we relax Eq. \eqref{fxs} and instead use
\begin{align}\label{rALDA_spin}
f^{rALDA}_{x,\sigma\sigma'}[n_\uparrow,n_\downarrow]=2f^{rALDA}_x[n]\delta_{\sigma\sigma'}+v^r[n]-v,
\end{align}
with $n=n_\sigma+n_{\sigma'}$. This satisfies Eq. \eqref{dyson_spin_condition} and ensures that the renormalized bare Coulomb interaction is replaced by the renormalized kernel, when we calculate the full Hartree-exchange kernel. The spin structure is then very similar to that of the ALDA Hartree exchange, the only difference being that we have replaced the bare Coulomb interaction with the renormalized Coulomb interaction and the ALDA kernel has been replaced by $\widetilde{f}^{rALDA}$, where both are evaluated on the total density. We note that this choice is by no means unique and in our previous work\cite{olsen_ralda1} we used the expression $f^{rALDA}_{x,\sigma\sigma'}=2f^{rALDA}_x[n_\sigma+n_{\sigma'}]\delta_{\sigma\sigma'}+v^r[n_\sigma+n_{\sigma'}]-v$. However we have found that Eq. \eqref{rALDA_spin} gives better results for atomization energies and we have used that approximation in the present work.

\subsubsection{Inhomogeneous systems}
It is straightforward to generalize the rALDA kernel Eq. \eqref{rALDA} to inhomogeneous systems by taking $r\rightarrow|\mathbf{r}-\mathbf{r}'|$ and $k_F\rightarrow(3\pi^2\tilde n(\mathbf{r},\mathbf{r}'))^{1/3}$. However, we are forced to introduce a two-point density $\tilde n(\mathbf{r},\mathbf{r}')$, which is not uniquely given by the present method. The only requirement is that $f_{xc}(\mathbf{r},\mathbf{r}')=f_{xc}(\mathbf{r}',\mathbf{r})$ which translates into $\tilde n(\mathbf{r},\mathbf{r}')=\tilde n(\mathbf{r}',\mathbf{r})$. In the following we will use two different flavors of the two-point density:
\begin{align}\label{density1}
\tilde n_1(\mathbf{r},\mathbf{r}')&=\frac{n(\mathbf{r})+n(\mathbf{r}')}{2},
\end{align}
and
\begin{align}\label{density2}
\tilde n_2(\mathbf{r},\mathbf{r}')&=n\Big(\frac{\mathbf{r}+\mathbf{r}'}{2}\Big).
\end{align}
The first of these choices is a simple average of the densities. In our view, this is the most natural choice since the two-point density should give the width of the kernel, if we regard this as a pure function of $|\mathbf{r}-\mathbf{r}'|$. Thus if $\mathbf{r}$ and $\mathbf{r}'$ belong to two well separated systems, the coupling originating from the kernel is determined by the width, which should be given by the average value of the densities at $\mathbf{r}$ and $\mathbf{r}'$. From the Dyson equation \eqref{dyson} it is clear that the Hartree-exchange-correlation kernel $f_{Hxc}(\mathbf{r},\mathbf{r}')$ provides a coupling between the non-interacting density response functions $\chi^{KS}(\mathbf{r}_1,\mathbf{r})$ and $\chi^{KS}(\mathbf{r}',\mathbf{r}_2)$ and the kernel thus naturally involves the densities in those points. In the following we will regard this as the physical choice and implicitly refer to this two-point density unless otherwise stated. In contrast, $\tilde n_2(\mathbf{r},\mathbf{r}')$ is the density at the average position, which becomes zero if $\mathbf{r}$ and $\mathbf{r}'$ belong to two well separated systems. This version of the two-point density is therefore not expected to describe van der Waals interactions correctly, but it has the great advantage that the kernel becomes a function of $r_-=|\mathbf{r}-\mathbf{r}'|$ and $\mathbf{r_+}=(\mathbf{r}+\mathbf{r}')/2$, which simplifies calculations in periodic systems. Nevertheless, \textit{a priori}, we would not expect $\tilde n_2(\mathbf{r},\mathbf{r}')$ to work well in very inhomogeneous systems.

\subsection{Plane wave implementation}
The renormalized ALDA functional has been implemented in the DFT code GPAW,\cite{mortensen, gpaw-paper} which uses the projector augmented wave (PAW) method.\cite{blochl} The response function is calculated in a plane wave basis set as described in Refs. [\onlinecite{jun,olsen_rpa2}]. We were not able to apply the analytic coupling constant integration Eq. \eqref{E_c_int} due to near singular behavior of $f_{Hx}^{rALDA}$. Instead we solve the Dyson equation \eqref{dyson} for 8 Gauss-Legendre lambda points and perform the coupling constant integration numerically. The frequency integration is performed using 16 Gauss-Legendre points with the highest point situated at 800 eV. 

The kernel Eq. \eqref{rALDA_exchange} with a general two-point density like Eq. \eqref{density1} is only invariant under simultaneous lattice translation in $\mathbf{r}$ and $\mathbf{r}'$. Its plane wave representation takes the form
\begin{align}\label{f_GG}
f^{rALDA}_{\mathbf{G}\mathbf{G}'}(\mathbf{q})&=\frac{1}{NV}\int_{NV}d\mathbf{r}\int_{NV}d\mathbf{r}'e^{-i(\mathbf{G+q})\cdot\mathbf{r}}\\
&\qquad\qquad\qquad\times f^{rALDA}_{x}(\mathbf{r},\mathbf{r}')e^{i(\mathbf{G'+q})\cdot\mathbf{r}'}\notag\\
&=\frac{1}{V}\int_{V}d\mathbf{r}\int_{V}d\mathbf{r}'e^{-i\mathbf{G}\cdot\mathbf{r}}f(\mathbf{q};\mathbf{r},\mathbf{r}')e^{i\mathbf{G}'\cdot\mathbf{r}'},\notag
\end{align}
where $\mathbf{G}$ and $\mathbf{G}'$ are reciprocal lattice vectors, $N$ is the number of sampled unit cells, $\mathbf{q}$ belongs to the first Brillouin zone, and
\begin{align}\label{f_tilde}
f(\mathbf{q};\mathbf{r},\mathbf{r}')=\frac{1}{N}\sum_{i,j}e^{i\mathbf{q}\cdot\mathbf{R}_{ij}}e^{-i\mathbf{q}\cdot(\mathbf{r}-\mathbf{r}')}f^{rALDA}_{x}(\mathbf{r},\mathbf{r}'+\mathbf{R}_{ij}).
\end{align}
Here we have introduced the lattice point difference $\mathbf{R}_{ij}=\mathbf{R}_i-\mathbf{R}_j$ and used that each of the $N$ sampled unit cell integrals in Eq. \eqref{f_GG} can be transferred into a single unit cell by letting $\mathbf{r}\rightarrow\mathbf{r}+\mathbf{R}_i$. We also used that $f^{rALDA}_{Hx}(\mathbf{r}+\mathbf{R}_j,\mathbf{r}'+\mathbf{R}_i)=f^{rALDA}_{Hx}(\mathbf{r},\mathbf{r}'+\mathbf{R}_{ij})$. The function $f(\mathbf{q};\mathbf{r},\mathbf{r}')$ is thus periodic in both $\mathbf{r}$ and $\mathbf{r}'$ and $f^{rALDA}_{\mathbf{G}\mathbf{G}'}(\mathbf{q})$ should be converged by sampling a sufficient number of lattice points. For isolated atoms and molecules, one should only use a single term in the sum, but make sure that the calculation is converged with respect to unit cell size. For solids, a consistent approach is to sample a set of unit cells, which matches the sampled $k$-point grid. Since, the sum in Eq. \eqref{f_tilde} only involves lattice point differences, a lot of terms are equal and we can reduce the double sum over lattice points to a sum over lattice point differences, where each term is weighted by the number of times that particular lattice point difference appears. %For example, an fcc structure with a $12\times12\times12$ lattice point sampling, gives rise to $(12^3)^2=2985984$ terms in the sum, which can be reduced to 12167 lattice point differences. 
Despite this reduction, the evaluation of the rALDA kernel for periodic systems still represents a major computational load, since we have to calculate the full two-point function $f^{rALDA}_{Hxc}(\mathbf{r},\mathbf{r}'+\mathbf{R}_{ij})$ for each lattice point difference.

\subsubsection{Average coordinates}
One way to circumvent the lattice point sampling for periodic systems, is to replace the general two-point density Eq. \eqref{density1} by the density at the average position Eq. \eqref{density2}. The rALDA kernel then becomes periodic in the average position and can be written $f^{rALDA}_{Hxc}[n(\mathbf{r_+})](\mathbf{r_-})$, where $\mathbf{r_-}=\mathbf{r}-\mathbf{r}'$ and $\mathbf{r_+}=(\mathbf{r}+\mathbf{r}')/2$. In the limit of $N\rightarrow\infty$ The Fourier transform in $N$ unit cells then becomes\footnote{In a finite volume, the transformation takes the form $\int_0^Ldx\int_0^Ldx'\rightarrow\int_{-L}^Ldx_-\int_0^{L-|x_-|}dx_+$, which we replace by $\int_{-L}^Ldx_-\int_0^Ldx_+$ for $L\rightarrow\infty$.}
\begin{align}
&f^{rALDA}_{x}(\mathbf{G},\mathbf{G}',\mathbf{q})=\frac{1}{NV}\int_{NV}d\mathbf{r}\int_{NV}d\mathbf{r}'\\
&\qquad\qquad\qquad\times e^{-i\mathbf{q}\cdot(\mathbf{r}-\mathbf{r}')}e^{-i\mathbf{G}\cdot\mathbf{r}}f^{rALDA}_{x}(\mathbf{r},\mathbf{r}')e^{i\mathbf{G}'\cdot\mathbf{r}'}\notag\\
&=\frac{1}{V}\int_{NV}d\mathbf{r_-}\int_{V}d\mathbf{r_+}\notag\\
&\qquad\qquad\qquad\times e^{-i\mathbf{q}\cdot\mathbf{r_-}}e^{-i\mathbf{G_+}\cdot\mathbf{r_+}}f^{rALDA}_{x}(\mathbf{r_-},\mathbf{r_+})e^{-i\mathbf{G_-}\cdot\mathbf{r_-}}\notag\\
&=\frac{1}{V}\int_{V}d\mathbf{r} e^{-i\mathbf{G_+}\cdot\mathbf{r}}f^{ALDA}_{x}[n(\mathbf{r})] \theta\Big(2k_F[n(\mathbf{r})]-|\mathbf{q}+\mathbf{G_-}|\Big)\notag,
\end{align}
where we defined $\mathbf{G_+}=\mathbf{G}-\mathbf{G}'$ and $\mathbf{G_-}=(\mathbf{G}+\mathbf{G}')/2$. In the last line we used the Fourier transform of the step function from Eq. \eqref{rALDA}.

The expression is thus very similar to the Fourier transform of the ALDA kernel except that it involves a density dependent step function.

\newpage

\section{Results}
The most striking improvement of the rALDA functional compared to RPA, is the accurate description of absolute correlation energies. The main motivation for the method is the accurate representation of the correlation energy of the HEG and in Ref. [\onlinecite{olsen_ralda1}], it was shown that this is also true for simple inhomogeneous systems. For example, the RPA correlation energy of a Hydrogen atom is -0.57 eV, whereas rALDA gives -0.02 eV. Note that this value differs from our previous work,\cite{olsen_ralda1} due to a different treatment of spin in the present work. Similarly for the H$_2$ molecule, RPA and rALDA yield absolute correlation energies of -2.2 eV and -1.2 eV respectively, which should be compared to the exact value of -1.1 eV.

We have previously demonstrated that the rALDA method also significantly improves the accuracy of molecular atomization energies compared to RPA.\cite{olsen_ralda1} In our previous work, the rALDA kernel was evaluated using the pseudo-density. In this work, the rALDA kernel is based on the all-electron density and is thus exactly represented within the PAW formalism.

In this section we will begin by stating the computational details. We then present results for atomization energies of molecules and show that the rALDA kernel accurately describes strong static correlation in the atomic limit of H$_2$ dissociation. Cohesive energies of solids are then discussed and the two methods presented in Eqs. \eqref{density1} and \eqref{density2} are compared. We then present C$_6$ coefficients of eight atoms evaluated with LDA, RPA and rALDA. Finally, we demonstrate that the rALDA method becomes very similar to RPA in the description of long range correlation, which we exemplify by the dissociation of a graphene bilayer and the binding energy of four molecular dimers.

\subsection{Computational details}
The calculation of RPA and rALDA correlation energies are performed in three steps. First a standard LDA calculation is carried out in a plane wave basis. The full plane wave Kohn-Sham Hamiltonian is diagonalized to obtain all unoccupied electronic states and eigenvalues. Finally, we choose a cutoff energy and calculate the Kohn-Sham response where we put the number of unoccupied bands equal to the number of plane waves defined by the cutoff and evaluate the correlation energy according to Eqs. \eqref{E_c} and \eqref{dyson}. The calculated correlation energies are added to non-selfconsistent Hartree-Fock energies evaluated on the same orbitals as the correlation energy. In general, we have found very little dependence on the input orbitals for Hartree-Fock energies.

All LDA and PBE calculations were performed with a 600 eV cutoff for the wavefunctions. For RPA and rALDA energies we use an additional cutoff, which in general is smaller than the wavefunction cutoff and defines the number of plane wave used to represent the response function. The calculations are challenging to converge with respect to this cutoff energy and for most calculations we have used an extrapolation scheme to obtain converged results. For sufficiently high cutoff energies the correlation energy is found to scale as\cite{harl08}
\begin{equation}
 E^c(E_{cut})=E^c+\frac{A}{E_{cut}^{3/2}}
\end{equation}
where $E^c$ is the converged result. For most considered systems the extrapolation was found to be very accurate when the cutoff is increased beyond 300 eV. Typically, the converged results are obtained by extrapolating correlation energies in the range 250-400 eV. When comparing systems with similar electronic structure in the same unit cell, energy differences converge much faster and in the case of a bilayer graphene we used a fixed cutoff of 200 eV to calculate the potential energy curve.

The rALDA kernel is evaluated on a real space grid where the grid spacing is set to $h=\pi/\sqrt{4E_{cut}}$. We have checked that rALDA results are converged with respect to this grid spacing which is always on the order $\sim0.16$ {\AA} when the wavefunction cutoff energy is 600 eV (see appendix A).

All calculations were performed on experimental geometries corrected for zero-point anharmonic effects. For the small atoms and molecules, the simulation was carried out in a periodic supercell where the shortest distance to an atom in a neighboring cell is 6 {\AA}.  For Si, Ge, Mg, Li, Al, Pd, Rh, Cu, Ag we used 8 {\AA} and for Na we used 10 {\AA}. All solids were simulated with a 12x12x12 gamma-centered $k$-point sampling for the correlation energy calculations and a Fermi-Dirac smearing of 0.01 eV. The Hartree-Fock energies were found to converge much slower with respect to $k$-point sampling and we typically used 18-22 $k$-points in each direction for this. For bilayer graphene we used a $k$-point sampling of 16x16 for both Hartree-Fock and correlation energies. 

\subsection{Cohesive energies of molecules and solids}

\subsubsection{Atomization energies of molecules}
\begin{table}[tb]
\begin{center}
\begin{tabular}{c|c|c||c|c|c|c|c|c}   
      & PBE & RPA@LDA & LDA & RPA@PBE & ALDA & rALDA &  Exp.\\
	\hline
H$_2$      & 105 & 109 & 113 & 109 & 110 & 111 & 109 \\
N$_2$      & 244 & 224 & 268 & 225 & 229 & 231 & 228 \\
O$_2$      & 144 & 112 & 174 & 103 & 155 & 118 & 120 \\
F$_2$      &  53 &  30 &  78 &  24 &  74 &  37 &  38 \\
CO         & 269 & 242 & 299 & 234 & 287 & 256 & 259 \\
HF         & 142 & 130 & 161 & 127 & 157 & 139 & 141 \\
H$_2$O     & 234 & 222 & 264 & 218 & 249 & 229 & 233 \\
C$_2$H$_2$ & 415 & 383 & 460 & 374 & 421 & 406 & 405 \\
CH$_4$     & 420 & 404 & 462 & 400 & 426 & 420 & 419 \\
NH$_3$     & 302 & 290 & 337 & 291 & 296 & 297 & 297 \\
\hline
MAE        & 8.7 & 10.3& 36.7& 14.4& 15.7& 1.9 &  \\
\end{tabular}
\end{center}
\caption{Atomization energies of diatomic molecules. The ALDA values are taken from Ref. [\onlinecite{furche_voorhis}] and experimental values (corrected for zero point vibrational energies) are taken from Ref. [\onlinecite{karton}]. All number are in kcal/mol. The bottom line shows the mean absolute error for this small test set.}
\label{tab:atomization}	
\end{table}

\begin{figure}[tb]
	\includegraphics[width=8.0 cm]{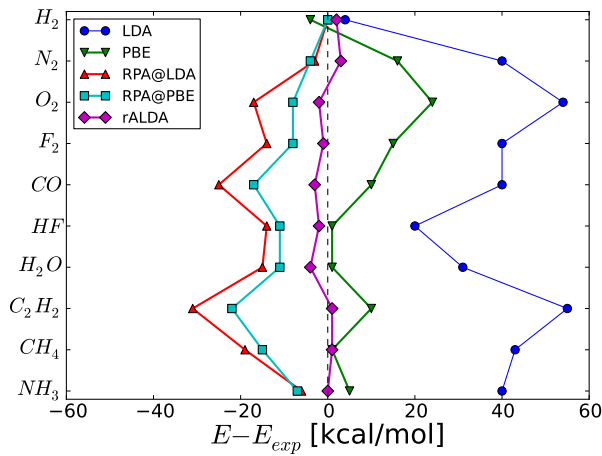} 
\caption{(color online). Atomization energies of 10 small molecules evaluated with LDA, PBE, RPA@PBE, RPA@LDA and rALDA.}
\label{fig:molecules}
\end{figure}

In table \ref{tab:atomization} we display the atomization energies of a small set of molecules evaluated with the rALDA functional and compare with LDA, PBE and RPA results. Some of the results differ slightly from previously published results\cite{olsen_ralda1} due to the use of all-electron densities instead of pseudo densities and due to a different treatment of spin-polarized systems. The accuracy of the rALDA is increased by a factor of 5 and 7 compared to RPA@PBE and RPA@LDA respectively and a factor of four compared to PBE. The deviations from experimental values are displayed in Fig. \ref{fig:molecules}. The PBE functional has a clear tendency to overestimate atomization energies, whereas RPA consistently underestimates atomization energies. The rALDA gives accurate results with no clear tendency to underbind or overbind. We also note that rALDA appears to be more accurate than RPA + SOSEX,\cite{gruneis, ren_review} which yields a MAE of 5.8 kcal/mol for small molecules. This value was obtained from the entire G2-1 test set, however, the MAE of RPA@PBE evaluated on G2-1 agree perfectly with the value stated in Tab. \ref{tab:atomization} and our small test set thus seems to give a representative value for the MAE of small molecules.

It is interesting to note that the performances of RPA@LDA is significantly worse than RPA@PBE. This is most likely due to a better description of KS eigenstates within PBE than within LDA. Thus, for a consistent comparison between RPA and rALDA, one should use RPA@LDA. When this is done, rALDA is then seen to improve the errors in atomization energies by nearly a factor of 7. Due to the bad performance of RPA@LDA it is tempting to speculate if rALDA would perform even better if evaluated on an improved set of KS eigenstates, for example those obtained with PBE. However, using an exchange correlation kernel, which is not derived from the orbitals on which it is applied is inconsistent and may yield numerical problems.\cite{furche_voorhis} In this case one would therefore need to apply a renormalized adiabatic PBE kernel in order to follow this path.

\subsubsection{Static correlation}
A surprising and appealing property of RPA is the good description of strong static correlation involved in the atomic limit of molecular dissociation for closed shell molecules.\cite{furche} However, within RPA molecular dissociation is correctly reproduced only if one corrects for the wrong RPA energy of the isolated atoms. This is due to the huge underestimation of the correlation energy in RPA. 

\begin{figure}[tb]
	\includegraphics[width=8.0 cm]{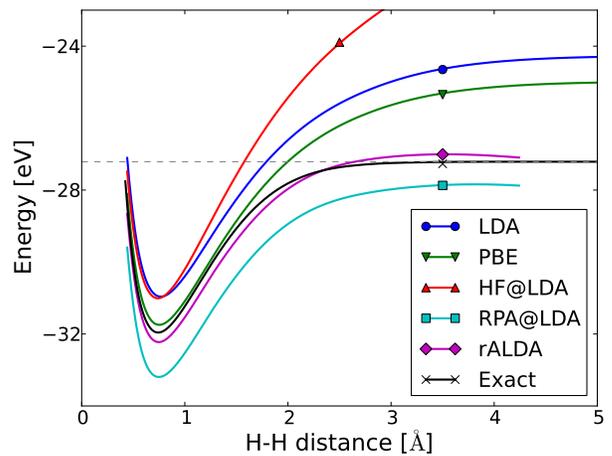} 
\caption{(color online). Dissociation curves of the H$_2$ molecule calculated with different functionals. The dashed line shows the energy of two isolated Hydrogen atoms (-1 Hartree). Each curve have been obtained by spline interpolation of 12 data points.}
\label{fig:H2}
\end{figure}

This error is largely eliminated in rALDA and the atomic limit of molecular dissociation is well reproduced in rALDA without the need for any corrections. This is shown in Fig. \ref{fig:H2} for the case of H$_2$. The rALDA dissociation curve approaches the rALDA energy of two isolated Hydrogen atoms (0.04 eV below one Hartree), whereas the RPA curve approached the RPA energy of two isolated Hydrogen atoms (1.2 eV below one Hartree). If these energies are used as references, the RPA and rALDA energy curves become practically indistinguishable.

It should be noted that like RPA, the rALDA energy curve exhibits a spurious maximum at $\sim 3.5$ {\AA}. Due to our plane wave implementation, simulations in large unit cells become prohibitly expensive and we were not able to calculate the energy all the way to the dissociation limit. The maximum is therefore only barely observable but distinct at closer inspection. In addition, from the present calculations it is not possible to conclude directly that the rALDA energy in the dissociation limit will approach the rALDA energy of two isolated hydrogen atoms. However, the long range correlation energy of rALDA approaches that of RPA in the dissociation limit and we thus expect the rALDA curve to coincide with the shifted RPA curve in this limit.

In contrast to the case of H$_2$, RPA fails dramatically in describing the dissociation of H$_2^+$. Again the situation is very similar for rALDA, except for the fact that the energy curve has been shifted by an amount corresponding to the RPA error in the correlation energy of a single H atom.

\subsubsection{Cohesive energies of solids}
\begin{figure}[tb]
	\includegraphics[width=8.0 cm]{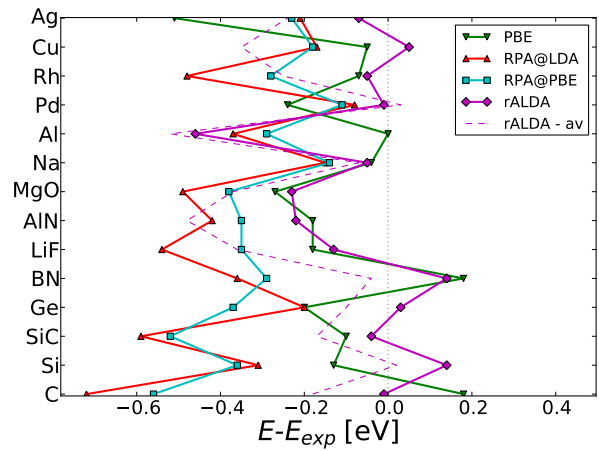} 
\caption{(color online). Cohesive energies of 14 solid evaluated with LDA, PBE, RPA@PBE, RPA@LDA and rALDA. The dashed line is rALDA results evaluated with the average density Eq. \eqref{density2}.}
\label{fig:solids}
\end{figure}

In Tab. \ref{tab:cohesive} we show the cohesive energies of 14 solids calculated with the LDA, PBE, RPA@LDA, RPA@PBE, and rALDA evaluated at experimental lattice constants. It is seen that PBE performs much better than RPA, which consistently underestimates the cohesive energies. For most solids, rALDA significantly improves the accuracy of RPA and becomes comparable to PBE. The only expecption is Al, where rALDA performs slightly worse than RPA. The overall absolute mean error of rALDA is 0.1 eV, which is a factor of 1.5 better than PBE and a factor of three better than RPA. The deviation from experimental values are plotted in Fig. \ref{fig:solids}. 

For the transition metals, we have found that it is very important to include the semicore states in the calculations. For Ag, Pd, and Rh we have thus included the entire $n=4$ shell in the calculations and for Cu we have included the 3s and 3p electrons as well as the 3d and 4s states. Both exchange and correlation energies are affected by this and treating the semicore states as a frozen core tend to reduce the atomization energies for these elements. For example, for RPA@PBE we find atomization energies of 3.48 and Tab. \ref{tab:cohesive}, where semicore states were included. The effect is less severe for the two noble transition metals, where the atomization energies increase by 0.1 eV when semicore states are included. It should be noted that the present RPA values are much closer to experimental values than previously published results \cite{harl10, olsen_rpa2} where semicore states were not included.

For periodic systems the rALDA method (with the physical two-point density \eqref{density1}) is significantly more computationally demanding than RPA calculations since the rALDA kernel has to be sampled in all unit cells corresponding to the $k$-point sampling. Furthermore, due to the long range of the Coulomb interaction it becomes difficult to converge the numerical Fourier transform of the Hartree-exchange kernel, in the sampled unit cells Eq. \eqref{f_GG}. For solids, we therefore perform the numerical Fourier transform of the pure exchange kernel, which has a much shorter range and then add the exact Fourier transform of the Hartree kernel $v_{\mathbf{G}}(\mathbf{q})=4\pi/|\mathbf{G}+\mathbf{q}|$.
\begin{equation}
 f^{Hx}_{\mathbf{G}\mathbf{G}'}(\mathbf{q})=\mathcal{F}_{\mathbf{G}\mathbf{G}'}\big[f^{x}(\mathbf{q},\mathbf{r},\mathbf{r}')\big]+v_{\mathbf{G}}(\mathbf{q})\delta_{\mathbf{G}\mathbf{G}'}.
\end{equation}
Here, $f^x$ is defined as in Eq. \eqref{f_tilde}. Since $f^x$ involves the bare Coulomb interaction, which diverges at $\mathbf{r}=\mathbf{r}'$, we represent this contribution by a spherical average over a small volume containing exactly one grid point. In appendix A, we present a convergence test, with respect to unit cell sampling for bulk Pd. 

In contrast, using the density at average coordinates \eqref{density2} allows one to to perform the calculations in a single unit cell, which makes the computational requirements similar to RPA. The results are shown as a dashed line in Fig. \ref{fig:solids} and are seen to be less accurate allthough the method still improves the RPA cohesive energies. 

\begin{table}[tb]
\begin{center}
\begin{tabular}{c|c|c||c|c|c|c}
    & PBE  & RPA@PBE  & LDA  & RPA@PBE  & rALDA & Expt. \\
\hline
C   & 7.73 & 6.99 & 8.94 & 6.83 & 7.54 & 7.55 \\
Si  & 4.55 & 4.32 & 5.63 & 4.37 & 4.82 & 4.68 \\
SiC & 6.38 & 5.96 & 7.38 & 5.89 & 6.44 & 6.48 \\
Ge  & 3.72 & 3.55 & 4.64 & 3.72 & 3.95 & 3.92 \\
BN  & 6.94 & 6.47 & 8.05 & 6.40 & 6.90 & 6.76 \\
LiF & 4.28 & 4.11 & 4.87 & 3.92 & 4.33 & 4.46 \\
AlN & 5.67 & 5.50 & 6.58 & 5.43 & 5.63 & 5.85 \\
MgO & 4.93 & 4.82 & 5.76 & 4.71 & 4.97 & 5.20 \\
Na  & 1.08 & 0.98 & 1.23 & 0.97 & 1.07 & 1.12 \\
Al  & 3.43 & 3.14 & 4.00 & 3.06 & 2.97 & 3.43 \\
Pd  & 3.70 & 3.83 & 5.07 & 3.86 & 3.93 & 3.94 \\
Rh  & 5.71 & 5.50 & 7.53 & 5.30 & 5.73 & 5.78 \\
Cu  & 3.47 & 3.34 & 4.49 & 3.35 & 3.57 & 3.52 \\
Ag  & 2.47 & 2.75 & 3.60 & 2.77 & 2.91 & 2.98 \\
\hline
MAE & 0.17 & 0.32 & 0.86 & 0.36 & 0.12 &  
\end{tabular}
\end{center}
\caption{Cohesive energy per atom of solids evaluated at the experimental lattice constant corrected for zero-point anharmonic effects. Experimental cohesive energies are corrected for zero point energy and taken from Ref. [\onlinecite{harl10}]. All numbers are in eV.}
\label{tab:cohesive}
\end{table}

Again, it is interesting to note that RPA@LDA performs worse than RPA@PBE. This could indicate that part of the errors obtained in LDA based calculations originate from a bad representation of the KS eigenstates and eigenvalues within LDA and we expect that this error is largest for the isolated atoms. For example, if we evaluate the Al atom with rALDA@PBE we obtain a correlation energy, that is 0.4 eV smaller than rALDA@LDA and this would put the cohesive energy of both Al and AlN right on top of the experimental value. In addition to the eigenstates and eigenvalues, the rALDA energies also depend explicitly on the density, which enters through the kernel. Thus, in rALDA an additional error may arise if the LDA density represents a poor approximation to the exact density.

\subsection{C$_6$ coefficients}
At large distances the dispersive interactions between atoms gives rise to a binding energy, which scales at $E_B(r)=C_6/r^6$. The C$_6$ coefficients depends on the polarizability of the atoms and can be calculated from the Casimir-Polder formula as
\begin{equation}
 C_6^{ij}=\frac{3}{\pi}\int_0^\infty\alpha_i(i\omega)\alpha_j(i\omega)d\omega,
\end{equation}
where  
\begin{equation}
 \alpha_i(i\omega)=-\int d\mathbf{r}d\mathbf{r'}z\chi_i(\mathbf{r},\mathbf{r'},i\omega)z'
\end{equation}
is the polarizability of atom $i$. The C$_6$ coefficients thus constitute a direct measure of the quality of a given approximation for the response function.

In Tab. \ref{tab:c6} we show the C$_6$ coefficients calculated for eight different atoms (with $i=j$) using LDA (Kohn-Sham response function), RPA@LDA, and rALDA. In a plane wave representation the polarizability involves a sum over the $G_z$ and $G_z'$ components of the response function and is hard to converge with respect to cutoff due to the discrete jumps in the number of $G_z$ with increasing cutoff. We were thus not able to converge the results to more than two significant digits, however, this is sufficient to assess the overall quality of the different approximations for $\chi$. 

The performance is highly dependent on the type of atoms considered. For the noble gas atoms LDA tends to overestimate the C$_6$ coefficients, whereas RPA and rALDA gives more accurate results. rALDA seems to take the lead in performance for the large noble gas atoms, but is inferior to RPA for He. For the alkaline elements Li and Na the Kohn-Sham LDA polarizability is very close to the exact value, whereas RPA significantly underestimates and rALDA performs much better. The alkaline earth species Be and Mg are severely overestimated by LDA and underestimated by RPA, whereas rALDA provides good agreement with the exact values. The overall performance is captured in the mean absolute relative error (MARE) and it is seen that rALDA outperforms LDA by a factor of 7 and RPA by a factor 3. However, due to the large scatter in result for different types of atoms, a proper assessment of the performance would require a larger set of atoms. Finally, the results seem to be very sensitive to the choice of input orbitals and RPA@HF has been shown to produce somewhat worse results for the noble gas atoms, but better results of the alkaline earth atoms.\cite{gould} Nevertheless, this small test system clearly indicates that rALDA is superior to RPA for C$_6$ coefficients.
\begin{table}[tb]
\begin{center}
\begin{tabular}{c|c|c|c|c}   
     & LDA & RPA@LDA & rALDA & Exact \\
	\hline
He   &  2.2 & 1.5 & 1.8  & 1.44 \\
Ne   &    9 &   6 &   7  & 6.48 \\
Ar   &  140 &  57 &  67  & 63.6 \\
Kr   &  280 & 110 & 130  & 130  \\
Li   & 1290 & 493 & 1180 & 1380 \\
Na   & 1520 & 560 & 1280 & 1470 \\
Be   &  590 & 163 &  243 & 219  \\
Mg   & 1400 & 370 &  570 & 630  \\
\hline
MARE & 0.79 & 0.29 & 0.11  & 
\end{tabular}
\end{center}
\caption{C$_6$ coefficients of eight atoms calculated with LDA, RPA, and rALDA. All values are in atomic units}
\label{tab:c6}	
\end{table}

\subsection{van der Waals interactions}
A major advantage of RPA, is the accurate representation of dispersive interactions, which are absent from any semi-local density functional. RPA has thus proven successful in describing the interlayer bonding in hexagonal Boron Nitride\cite{marini} and graphite\cite{lebegue} as well as the interaction between graphene and metal surfaces\cite{olsen_rpa1, olsen_rpa2, mittendorfer}. Furthermore, RPA has been shown to yield a three-fold improvement in binding energies of the S22 test set\cite{s22} of weakly bound molecular dimers, compared to PBE.\cite{ren_review} 

Since the main merit of RPA is the applicability to van der Waals bonded systems, it is of vital importance that any extension of RPA does not destroy the good description of long range correlation. The rALDA kernel in Eq. \eqref{rALDA} approaches the Hartree kernel for $r\rightarrow\infty$ and we therefore expect rALDA to produce results similar to RPA for van der Waals bonded systems. Below we will explicitly verify that this is the case by calculating binding energies for a few members of the S22 test set and the potential energy curve of bilayer graphene.

\subsubsection{Binding energies of molecular dimers}
The binding energies of four members of the S22 test set are displayed in Tab. \ref{tab:s22}. The rALDA energies are seen to be very similar to RPA binding energies. Except for the case of the CH$_4$ dimer, the bonding in these particular dimers does not have a purely non-local character and PBE actually performs better than RPA and rALDA. We also show the energies obtained with the van der Waals density function of Ref. \onlinecite{dion}, which is seen to perform worse than RPA/rALDA for the relatively strongly bound (H$_2$CO$_2$)$_2$ and similar to RPA/rALDA for the remaining three dimers.

Our plane wave wave implementation is ill suited for calculations of this type, which require large supercells and the RPA and rALDA calculations may not be completely converged with respect to supercell size. However, the example does illustrate the similarity of RPA and rALDA for dispersive interactions between molecules.
\begin{table}[t]
\begin{center}
\begin{tabular}{c|c||c|c||c|c|c|c}   
      & vdW-DF & PBE & RPA & LDA & RPA & rALDA & Ref. \\
	\hline
(H$_2$CO$_2$)$_2$ & 0.64 & 0.77 & 0.71 & 1.16 & 0.70 & 0.71 & 0.82 \\
(H$_2$O)$_2$      & 0.17 & 0.21 & 0.18 & 0.34 & 0.17 & 0.17 & 0.22 \\
(NH$_3$)$_2$      & 0.10 & 0.12 & 0.12 & 0.22 & 0.10 & 0.10 & 0.14 \\
(CH$_4$)$_2$      & 0.03 & 0.00 & 0.02 & 0.04 & 0.01 & 0.01 & 0.02
\end{tabular}
\end{center}
\caption{Dimer binding energies for four members of the S22 data set. The reference energies and applied geometries are based on coupled cluster (CCSD(T)) calculations from Ref. \onlinecite{s22}. All number are in eV.}
\label{tab:s22}
\end{table}

\subsubsection{Bilayer graphene}
\begin{figure}[tb]
	\includegraphics[width=8.0 cm]{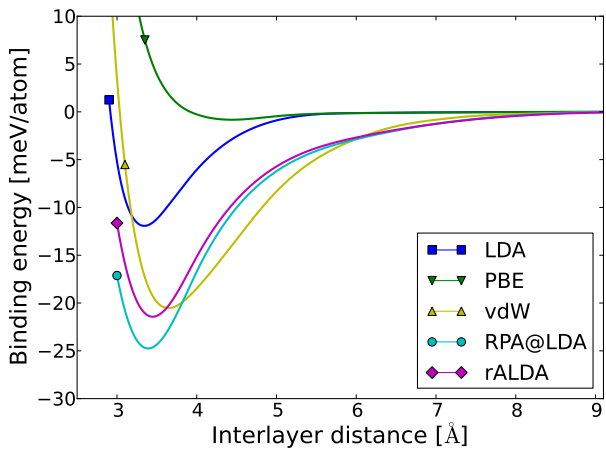}  
\caption{(color online). Potential energy curve for bilayer graphene calculated with different functionals. Each curve has been obtained by spline interpolation of 7 data points.}
\label{fig:bilayer}
\end{figure}

In Fig. \ref{fig:bilayer} we show the binding energy curve of an A-B stacked bilayer of graphene. As in the case of graphite, the PBE functional predicts a very weak binding energy and an equilibrium distance of 4.44 {\AA}. The van der Waals density functional of Ref. [\onlinecite{dion}] gives a binding energy of 22 meV and an equilibrium distance of 3.65 {\AA}. RPA gives a binding energy of 25 meV and an equilibrium distance of 3.39 {\AA}, whereas rALDA gives a binding energy of 22 meV and an equilibrium distance of 3.45 {\AA}. We should note that the binding energy curves were obtained by fitting a rather rough set of interlayer distances and more accurate results would require more accurate sampling of the binding energy curve. The RPA and rALDA curves coincide a large distances and decay slower than the tail of the curve obtained with the van der Waals functional.

In general RPA has a tendency to underbind and the reduced binding energy of rALDA compared to RPA could imply that rALDA actually performs worse than RPA in this case. However, in the case of graphite, RPA seems to produce the exact binding energy between layers\cite{lebegue} and it is not clear if RPA is also expected to underbind in this case. Furthermore, RPA and rALDA seems to give identical results for the few S22 dimers and while we do not have accurate experimental data for the binding energy and equilibrium distance of bilayer graphene, Fig. \ref{fig:bilayer} merely serves to illustrate the applicability and similarity of RPA and rALDA for the description of long range dispersive interactions.

\section{Summary and outlook}
We have assessed the performance of a non-local adiabatic kernel (rALDA) for electronic correlation energies. For atomization energies of small molecules and cohesive energies of insulators and metals, the kernel performs significantly better than RPA. We have found inclusion of semi-core states is crucial for the description of transition metals at the RPA and rALDA level. For the small molecules, we also obtain better agreement with experiments than the SOSEX extension to RPA. The method preserves (but does not improve) the good description of static correlation and dispersive interactions, which is the main merit of RPA. The kernel also improves the description of C$_6$ coefficients compared to RPA.

In our opinion, a major advantage of the present approach is the unique choice of input orbitals and eigenvalues, which should match the adiabatic (renormalized) kernel. While LDA orbitals may not be the most accurate choice for evaluating the Kohn-Sham response function, it is satisfactory that we do not have to make an arbitrary choice, which can have a severe influence on the results. From the point of view of TDDFT, RPA corresponds to the time-dependent Hartree approximation and a consistent choice in that case would be to adopt Hartree orbitals and eigenvalues. However, this would be a very bad starting point and RPA calculations are usually performed on top of semilocal DFT orbitals or Hartree-Fock orbitals. In this sense, the non-selfconsistent rALDA calculations represent a more complete and consistent scheme than non-selfconsistent RPA.

The computational cost of the method is larger than RPA, but certainly cheaper than time-dependent exact exchange or SOSEX. The main differences compared to RPA is the evaluation of the renormalized non-local kernel. For non-periodic systems this evaluation does not comprise a major extension of the computational time compared to RPA. For the cohesive energies of solids a large unit cell sampling is required to converge results and the calculations are somewhat more time-consuming than RPA. On the other hand for bilayer graphene, the potential energy surfaces are converged with a unit cell sampling of 4 adjacent cell and the computational cost of rALDA is just 20 {\%} extra compared to RPA. The cohesive energies of solids are important for an assessment of the method, but for future application to larger systems (for example molecular adsorption or reaction barriers at metal surfaces) the calculations will be dominated by the evaluation of the repsonse function and the computational cost becomes comparable to RPA. Compared to RPA, another complication is the spin-dependence of the kernel, which forces us to solve the full spin-dependent Dyson equation instead of working with spin-summed quantities. In our plane wave implementation, this leads to memory problems for spin-polarized systems when considering large unit cells and high cutoff energies. Finally, the coupling constant integration is carried out numerically due to near-singular behavior of the kernel. However, this problem is very likely to originate from the plane-wave representation of the kernel and a different implementation could probably solve this problem and apply the analytic result \eqref{E_c_int}.

In the present paper we have only discussed the renormalized ALDA kernel. However, it should be straightforward to generalize this to a renormalized adiabatic PBE kernel or in fact, any semi-local exchange kernel. We would expect this to improve results further due to more accurate initial orbitals and eigenvalues and because the kernel would then contain gradient corrections, which are likely to improve the description of the interacting response function. The cutoff scheme thus implies an entire hierarchy of renormalized adiabatic kernels for correlation energy calculations. One caveat, is the fact that many semi-local approximations to the exchange correlation energy work well due to error cancellation between exchange and correlation and it is possible that one would have to include the correlation part of the kernel as well, in order to obtain highly accurate results. This would, however, be straightforward and one would simply have to evaluate the full kernel explicitly along the adiabatic connection since the linear scaling of pure exchange is lost. It would also be very interesting to compare the present results, with those obtained with the fitted non-local kernel of Ref. \cite{jung}, which is defined by a similar cuf off procedure. We will leave these issues to future work.

\appendix
\section{Convergence tests}
\subsection{Grid spacing}
\begin{figure}[tb]
	\includegraphics[width=7.0 cm]{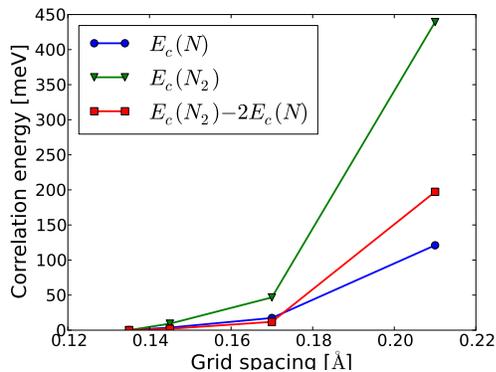}  
\caption{(color online). Convergence of rALDA correlation energy with respect to grid spacing for the atomization energy of N$_2$.}
\label{fig:N2_grid}
\end{figure}

The two-point kernel Eq. \eqref{f_GG} is evaluated using a density, which is represented on a real space grid. We have not been able to derive an explicit PAW correction for the kernel, and have used the all-electron density for calculations in this work. Since the all-electron density varies rapidly in the core region of atoms, it is not obvious that it is possible to converge the kernel with respect to grid spacing. However, the ALDA exchange kernel approaches zero for large densities and as it turns out, the rapidly oscillating core region does not contribute much to the rALDA kernel. This is illustarted in Fig. \ref{fig:N2_grid} where the correlation energy of an N$_2$ molecule is plotted for decreasing grid spacing. The energy difference (contribution to the atomization energy) converges rapidly and is accurate to within 10 meV at 0.17 {\AA}. This is the value used for all calculations in the present work and is close to the default value of 0.18 {\AA} in GPAW. The correlation energy of a single N atom is also close to convergence at this value, whereas a slightly smaller grid spacing is required for N$_2$.

\subsection{Unit cells}
For the molecular and atomic systems in this work we have obtained the plane wave representation of the rALDA kernel by evaluating the full Hartree-exchange kernel in real space and performing a numerical Fourier transform
\begin{equation}\label{rALDA1}
f^{Hx}_{\mathbf{G}\mathbf{G}'}(\mathbf{q=0})=\mathcal{F}_{\mathbf{G}\mathbf{G'}}\big[\widetilde f^{rALDA}_x[n](\mathbf{r},\mathbf{r}')+v^r[n](\mathbf{r},\mathbf{r}')\big].
\end{equation}
in a single unit cell. This results in a truncation on the kernel whenever $\mathbf{r}$ or $\mathbf{r}'$ is outside the unit cell, which means that the system will not interact with its periodic images. Another major advantage of this approach is that the full Hartree-exchange kernel is finite for $\mathbf{r}=\mathbf{r}'$ Eq. \eqref{kernel_origin}, which is not true for the bare exchange kernel Eq. \eqref{rALDA_exchange}.

\begin{figure}[tb]
	\includegraphics[width=4.0 cm]{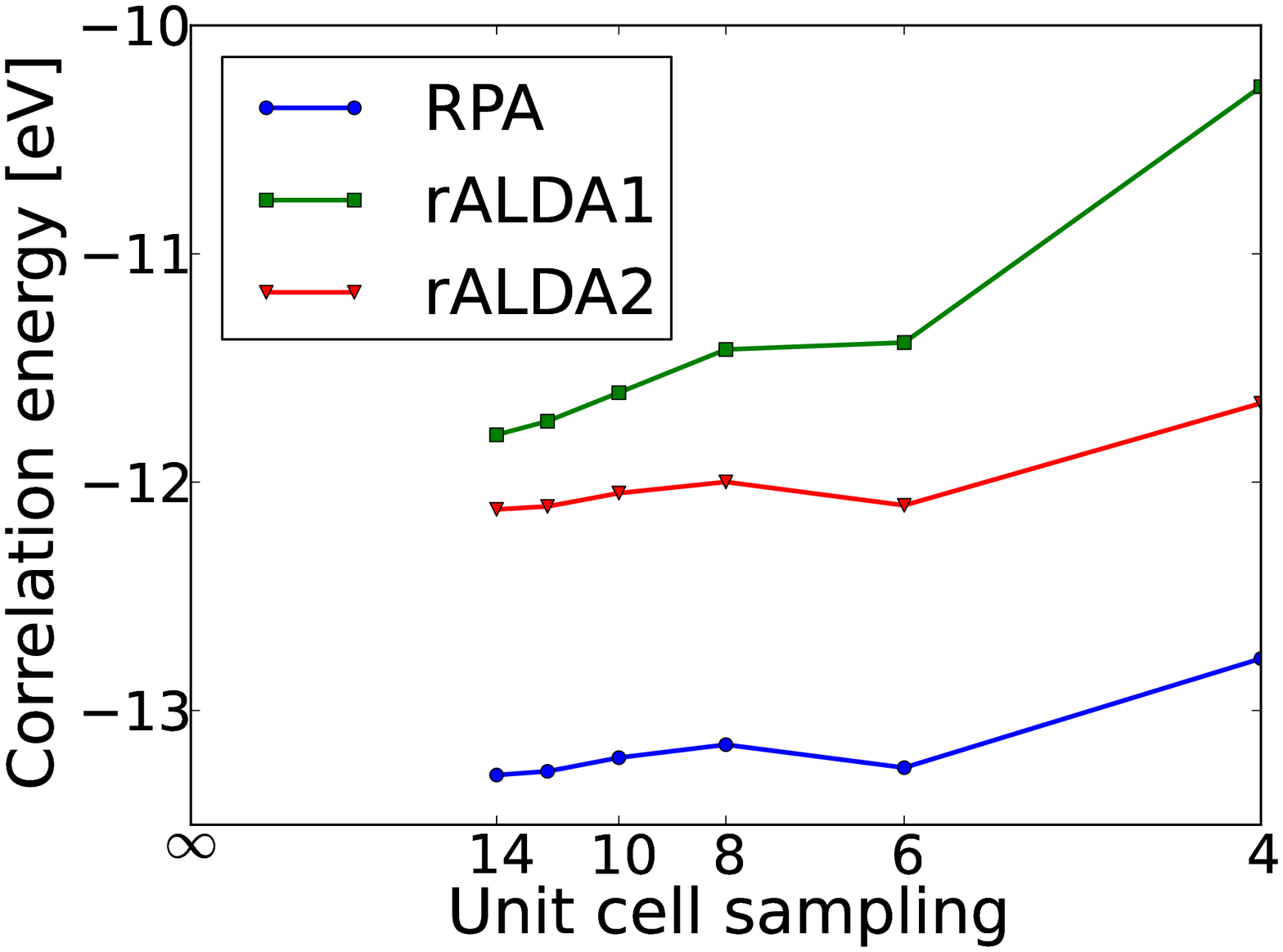}  
	\includegraphics[width=4.0 cm]{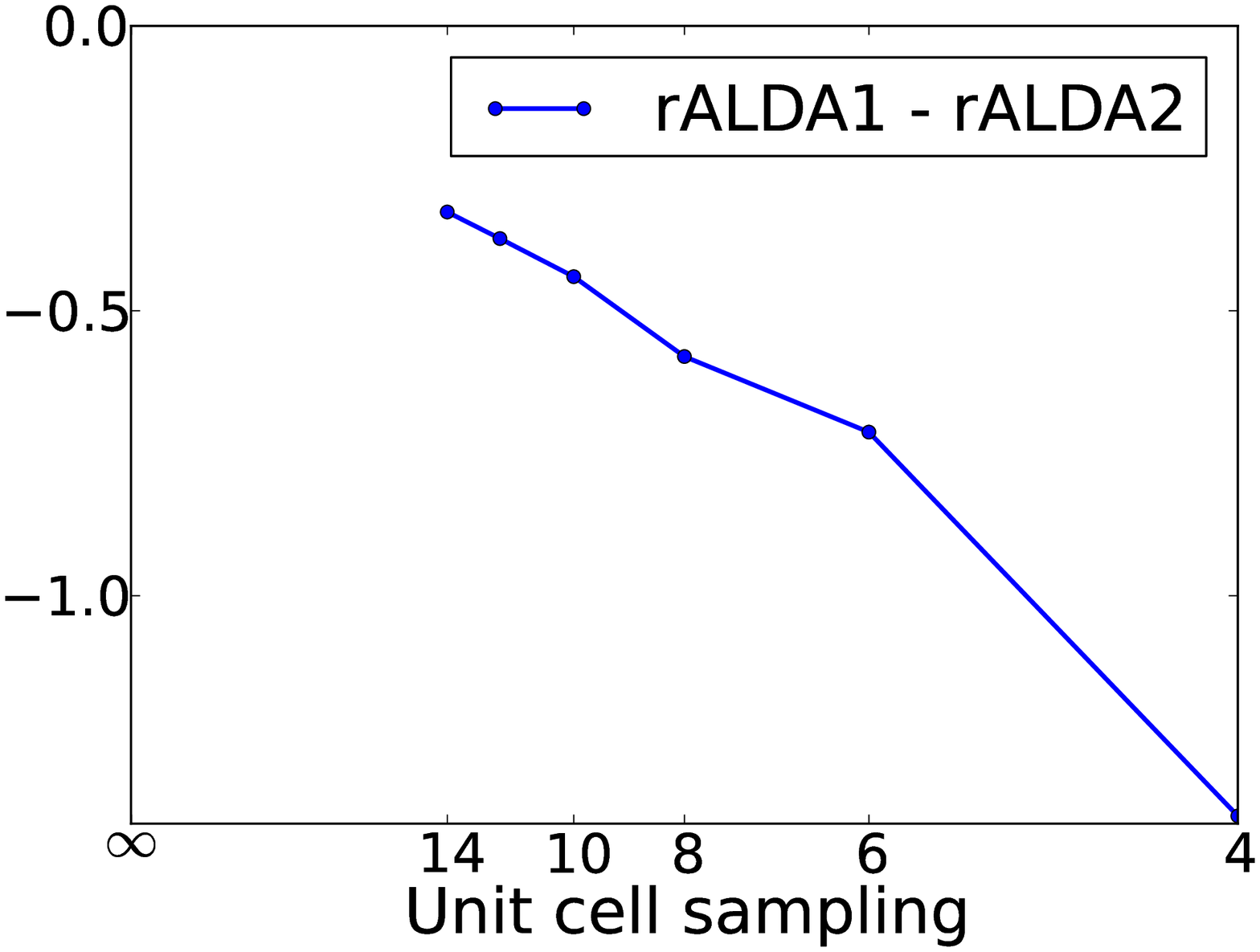}  
\caption{(color online). Convergence of rALDA correlation energy with respect to sampled unit cells of solid Pd. rALDA1 denotes the implementation Eq. \eqref{rALDA1} and rALDA2 is the implementation Eq. \eqref{rALDA2}. The scale on the principle axes is $1/N$.} 
\label{fig:Pd_kpts}
\end{figure}

For solids, however, it is important to take into account the long range nature of the Coulomb interaction. The renormalized Coulomb interaction $v^r$ approaches the bare Coulomb interaction when $r\rightarrow\infty$ and it becomes difficult to converge the Fourier transform of this with respect to the number of sampled unit cells. For periodic systems, we therefore only represent the exchange part of the kernel in real space and add the exact Fourier transform of the Hartree kernel:
\begin{align}\label{rALDA2}
f^{Hx}_{\mathbf{G}\mathbf{G}'}(\mathbf{q})&=v_{\mathbf{G}}(\mathbf{q})\delta_{\mathbf{G}\mathbf{G}'}+
\mathcal{F}_{\mathbf{G+q}\mathbf{G'+q}}\big[f^{rALDA}_x(\mathbf{r},\mathbf{r}')\big],
\end{align}
where $v_{\mathbf{G}}(\mathbf{q})=4\pi/|\mathbf{G+q}|^2$. This representation is easier to converge since $f^{rALDA}_x(r)\rightarrow\sin(2k_F[n]r)/r$ for $r\rightarrow\infty$. This is illustrated in Fig. \ref{fig:Pd_kpts} where we compare the convergence of the two implementations with respect to the number of sampled unit cells. It is seen that Eq. \eqref{rALDA2} exhibits convergence behavior very similar to RPA, which means that the convergence is largely governed by Brillouin zone sampling of Kohn-Sham states. In contrast, the implementation Eq. \eqref{rALDA1} converges as $\sim1/N$ where $N$ is the number of sampled unit cells. We also show the difference between the two methods, which slowly approaches zero in the limit of $N\rightarrow\infty$. However, the bare exchange kernel diverges for $\mathbf{r}=\mathbf{r}'$, and this divergence is represented by a spherical average around a single grid point, which make the calculations converge slightly slower with respect to grid spacing.

\begin{acknowledgments}
The authors acknowledge support from the Danish Council for Independent Research's Sapere Audie Program, grant no 11-1051390. The Center for
Nanostructured Graphene is sponsored by the Danish National Research Foundation.
\end{acknowledgments}

%\bibliography{bibfile}{}

\end{document}